\documentclass[trackchanges,twocolumn]{aastex701}
\usepackage{booktabs}
\usepackage{amsmath}

\begin{document}

\title{First Discovery of a Hot Subdwarf B Variable Star in a Young Open Cluster}

\correspondingauthor{Huahui Yan, Xiangcun Meng}

\author[orcid=0009-0005-8565-0858]{Huahui Yan}
\affiliation{School of Physics and Astronomy, Sun Yat-sen University, DaXue Road, Zhuhai 519082, GuangDong, People’s Republic of China.}
\affiliation{CSST Science Center for the Guangdong-Hongkong-Macau Greater Bay Area, Sun Yat-sen University, DaXue Road, Zhuhai 519082, GuangDong, People’s Republic of China}
\email[show]{yanhh8@mail2.sysu.edu.cn}

\author{Ruijie He} 
\affiliation{International Centre of Supernovae (ICESUN), Yunnan Key Laboratory of Supernova Research, Yunnan Observatories, Chinese Academy of Sciences (CAS), Kunming 650216, China}
\email[]{}

\author[0000-0002-4591-1903]{David R. Miller}
\affiliation{Department of Physics and Astronomy, University of British Columbia, Vancouver, BC V6T 1Z1, Canada}
\email{drmiller@phas.ubc.ca}

\author{Shunyi Lan} 
\affiliation{International Centre of Supernovae (ICESUN), Yunnan Key Laboratory of Supernova Research, Yunnan Observatories, Chinese Academy of Sciences (CAS), Kunming 650216, China}
\email[]{}

\author{Xiaodian Chen} 
\affiliation{National Astronomical Observatories, Chinese Academy of Sciences, Beijing 100101, China}
\affiliation{Institute for Frontiers in Astronomy and Astrophysics, Beijing Normal University, Beijing 102206, China}
\email[]{}

\author{Yang Huang} 
\affiliation{School of Astronomy and Space Science, University of Chinese Academy of Sciences, Beijing 100049, China}
\affiliation{National Astronomical Observatories, Chinese Academy of Sciences, Beijing 100101, China}
\email[]{}

\author{Jingkun Zhao} 
\affiliation{CAS Key Laboratory of Optical Astronomy, National Astronomical Observatories, Chinese Academy of Sciences, Beijing 100101, People’s Republic of China}
\email[]{}

\author{Peter Nemeth} 
\affiliation{Astroserver.org, F\H{o} t\'{e}r 1, 8533 Malomsok, Hungary }
\email[]{}

\author{Chuanjie Zheng} 
\affiliation{School of Astronomy and Space Science, University of Chinese Academy of Sciences, Beijing 100049, China}
\email[]{}

\author{Yingkang Sun} 
\affiliation{National Astronomical Observatories, Chinese Academy of Sciences, Beijing 100101, China}
\email[]{}

\author{Xiangcun Meng} 
\affiliation{International Centre of Supernovae (ICESUN), Yunnan Key Laboratory of Supernova Research, Yunnan Observatories, Chinese Academy of Sciences (CAS), Kunming 650216, China}
\email[show]{xiangcunmeng@ynao.ac.cn}

\author{Chengyuan Li} 
\affiliation{School of Physics and Astronomy, Sun Yat-sen University, DaXue Road, Zhuhai 519082, GuangDong, People’s Republic of China.}
\affiliation{CSST Science Center for the Guangdong-Hongkong-Macau Greater Bay Area, Sun Yat-sen University, DaXue Road, Zhuhai 519082, GuangDong, People’s Republic of China}
\email{}

\begin{abstract}
Hot subdwarf B (sdB) stars are core helium-burning objects on the extreme horizontal branch, yet their formation histories remain uncertain because field sdBs generally lack reliable ages. Cluster membership provides an independent age constraint, but observational constraints on sdB formation in young clusters remain limited. Here we report the first spectroscopically confirmed sdB star in a young open cluster (OC): G302 in NGC~2437. A systematic search of OCs using Gaia DR3 astrometry identifies G302 as a high-confidence member, supported by its consistent radial velocity (RV). Isochrone fitting yields a cluster age of $\sim380$~Myr. Zwicky Transient Facility (ZTF) photometry reveals large-amplitude pulsations with a dominant period of $6.263$~min, while spectroscopy yields $T_{\rm eff}=28\,317\pm708$~K and $\log g=5.33\pm0.10$.
Using the cluster age as a constraint, we fit Modules for Experiments in Stellar Astrophysics (\texttt{MESA}) models to the atmospheric parameters and estimate the progenitor mass of an sdB descended from an intermediate-mass star. In models where the envelope is stripped during the Hertzsprung gap, a $2.8$--$3.1\,M_\odot$ progenitor (best fit $2.9\,M_\odot$) ignites helium non-degenerately and evolves into a $0.41$--$0.45\,M_\odot$ sdB consistent with the observed parameters. The young cluster age, together with published binary population synthesis predictions, suggests that G302 most likely formed through
the first stable Roche-lobe overflow (RLOF) channel. G302 provides an age-calibrated benchmark for testing sdB formation from intermediate-mass progenitors.
\end{abstract}


\section{Introduction} 

Hot subdwarfs (sdOs and sdBs) are evolved  stars with O- or B-type spectra located on or near the extreme horizontal branch (EHB), with luminosities between the main sequence (MS) and the white dwarf (WD) sequence \citep{Heber2016}. Most of them are core helium-burning stars of about half a solar mass, surrounded by a hydrogen envelope too thin ($\lesssim 0.02\,M_\odot$) to sustain hydrogen-shell burning \citep{Heber2016,Geier2017}. To reach this configuration, the progenitor must lose nearly all of its hydrogen envelope during post-main-sequence evolution while retaining a helium core capable of subsequent helium ignition. Such fine-tuned mass loss is difficult to achieve in single-star evolution, so binary interactions are generally invoked. The binary population synthesis (BPS) models of \citet{Han2002,Han2003} identified three main groups of formation channels: stable Roche-lobe overflow (RLOF), which leaves the sdB in a wide binary; common-envelope (CE) ejection, which produces a close binary with an MS or WD companion; and the merger of two helium WDs, which produces a single sdB. These models reproduce the main observed properties of the hot subdwarf population, and the high fraction of binaries among H-rich hot subdwarfs supports the binary picture \citep{Maxted2001,Stark2003,He2025AA,He2025MNRAS}. About 10\% of hot subdwarfs have helium-dominated atmospheres (He-sdB and He-sdO; \citealt{Geier2017}). Because most of them are apparently single, they are usually attributed to mergers, either of two helium WDs or of a helium WD with a low-mass MS star \citep{Zhang2012,Justham2011}.

Despite this progress, the relative importance of the different channels, and the progenitors that feed each of them, remain poorly constrained by observations. The BPS models predict that the channel contributions depend strongly on the age of the stellar population \citep{Han2008}. In young populations ($\lesssim 1$\,Gyr), the first stable RLOF channel is expected to dominate, and the progenitors are intermediate-mass stars ($\gtrsim 2\,M_\odot$). These progenitors ignite helium in non-degenerate cores, so the resulting sdBs can have masses well below the canonical $\sim 0.47\,M_\odot$, down to $\sim 0.3\,M_\odot$ \citep{Han2002}. In older populations, CE ejection and helium-WD mergers become increasingly important. The same diversity of formation pathways appears in the related class of blue large-amplitude pulsators (BLAPs; \citealt{Pietrukowicz2017,Kupfer2019}), whose origin is still debated. Proposed scenarios include low-mass stripped stars produced by binary interaction \citep{Byrne2018}, and surviving companions of Type Ia supernovae \citep{Meng2020}. \citet{Meng2021} further linked BLAPs to hot subdwarfs of intermediate helium abundance. Testing these predictions requires knowing the age of the population to which an object belongs. For field hot subdwarfs, however, the age and hence the initial mass of the progenitor cannot be determined. As a result, the population-age dependence of the formation channels, which is a key prediction of the BPS models, has never been tested directly for individual objects.

Star clusters offer a way around this problem. Stars in a cluster form from the same molecular cloud and therefore share a common age and metallicity. A hot subdwarf confirmed as a cluster member inherits these properties, which directly constrains the initial mass of its progenitor and the time available for its evolution. Many hot subdwarfs have been identified in star clusters, such as NGC~6752 \citep{MoniBidin2006}, $\omega$~Centauri (NGC~5139) \citep{Moehler2007,Latour2018}, NGC~188 \citep{Landsman2003,Rani2021}, and NGC~6791 \citep{Kaluzny1992,Liebert1994,Sanjayan2022}.  These examples are found in old clusters, where hot-subdwarf formation mainly probes low-mass progenitors that develop degenerate helium cores. Young clusters provide complementary opportunities to study envelope stripping and non-degenerate helium ignition in intermediate-mass stars. An important precedent is the Be+sdO binary $\phi$~Persei, whose hot companion has been spectroscopically confirmed \citep{Gies1998} and whose distance and kinematics support a likely association with the young $\alpha$~Per cluster \citep{Mourard2015}. Its sdO companion has a mass of approximately $1.2 \,M_\odot$ \citep{Mourard2015} and has been interpreted as a helium-shell-burning star \citep{Schootemeijer2018}, distinguishing it from low-mass, core-helium-burning sdB stars.

More recently, \citet{Nedhath2025} used multi-band ultraviolet photometry and spectral energy distribution (SED) fitting to identify 16 candidate sdOB-type stripped companions to Be stars in the young open cluster NGC~663. These photometric identifications require spectroscopic follow-up to confirm the nature of the hot companions. Spectroscopically characterized low-mass sdB stars in young clusters would provide complementary constraints on stable Roche-lobe overflow and non-degenerate helium ignition in intermediate-mass progenitors.

In this work, we carry out a systematic search for hot subdwarfs in Galactic OCs using Gaia DR3 \citep{GaiaCollaboration2023A&A...674A...1G} astrometry. We identify a spectroscopically confirmed sdB, G302, as a member of the young OC NGC~2437 ($\sim 380$\,Myr). Using the cluster age as a constraint, we then use stellar evolution models to infer the progenitor of G302 and its likely formation channel.

This paper is organized as follows. Section~\ref{sec:membership} establishes the cluster membership of G302. Section~\ref{sec:params} presents its photometric and spectroscopic parameters and the cluster age. Section~\ref{sec:evolution} discusses its evolutionary channel using MESA code. We summarize our conclusions in Section~\ref{sec:summary}.

\section{Cluster Membership Determination} \label{sec:membership}

\subsection{Sample}
Our OC sample is derived from the catalogue of \citet{Hunt2023,Hunt2024}, which builds upon an initial blind all-sky search of \textit{Gaia} DR3 data using the HDBSCAN algorithm. To ensure high reliability and exclude unbound density fluctuations or chance alignments, we applied their recommended high-probability selection criteria: a CMD Class $> 0.5$ and a Clustering Significance Test (CST) $> 5\sigma$. This robust filtering restricts our analysis to  likely gravitationally bound systems, resulting in a final sample of 3,530 OCs adopted for this work. For the hot subdwarf stars, we utilized two comprehensive catalogues compiled by \citet{Culpa2022} based on Gaia EDR3 data. The first dataset is the \textit{Known Hot Subdwarf Catalogue}, consisting of 6,616 spectroscopically confirmed sources from the literature and recent large-scale surveys. To extend the sample volume, we also adopted their \textit{Candidate Catalogue}, which comprises 61,585 sources. These candidates were identified through a robust dual-selection strategy: a `Parallax Selection' targeting sources with reliable astrometry ($\varpi/\sigma_\varpi > 5$), and a `Reduced Proper Motion Selection' utilized as a luminosity proxy for sources with larger parallax errors. This dual approach substantially improves sample completeness while filtering out contaminants such as MS stars and WDs, although significant incompleteness likely remains, particularly for the reduced proper-motion sample.

\subsection{Method}

\begin{figure*}[ht!]
	\centering
	\includegraphics[width=\textwidth]{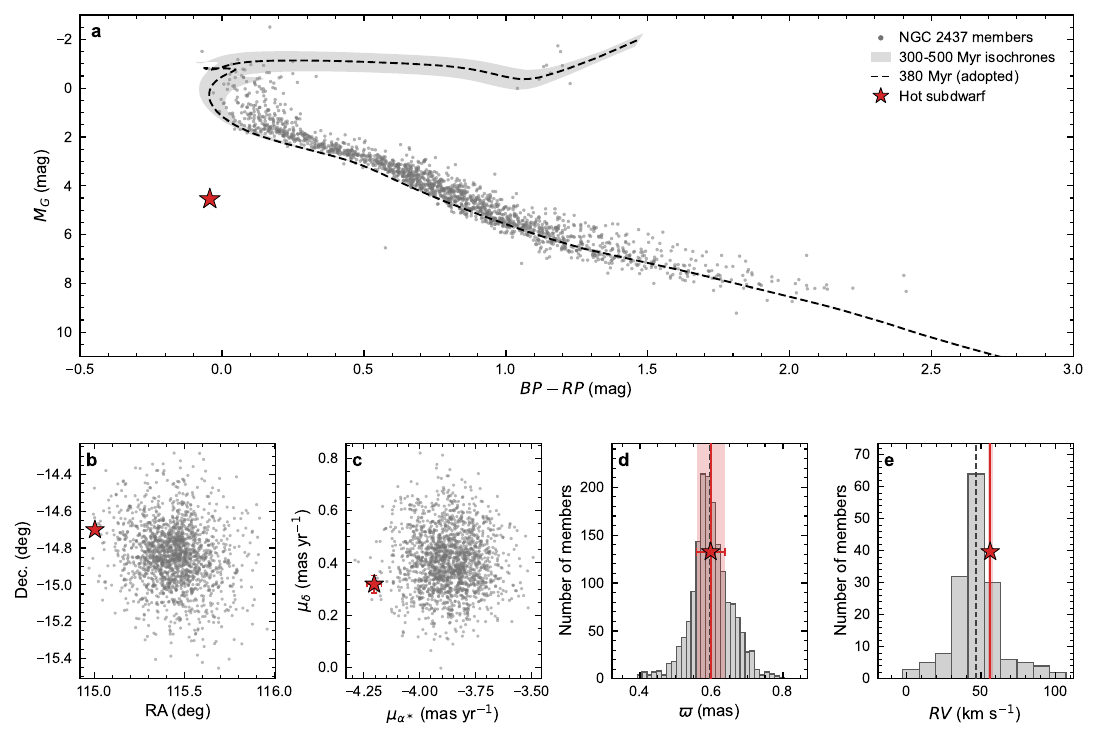}
\caption{
	Cluster membership of the hot subdwarf G302 in NGC~2437.
	(a) Gaia color--magnitude diagram of the cluster. Grey dots are the cluster members from \cite{Hunt2023}, and the red star marks G302. The grey shaded band spans the isochrones with ages of $300$--$500\,{\rm Myr}$, and the dashed curve is the $380\,{\rm Myr}$ isochrone adopted in this work.
	Panels (b)--(e) show the spatial distribution, the proper-motion distribution ($\mu_{\alpha*}$ versus $\mu_\delta$), the parallax ($\varpi$) distribution, and the radial velocity (RV) distribution of the cluster members, respectively, with the RV histogram including only members with available RV measurements.
	In panels (d) and (e), the grey dashed line marks the median value of the cluster members. The red solid line and shaded band show the value of G302 and its $1\sigma$ uncertainty.
	\label{fig1}}
\end{figure*}

To identify hot subdwarf stars associated with OCs, we implemented a two-stage selection strategy: a broad initial screening followed by a rigorous statistical evaluation. First, we established a preliminary sample by applying broad spatial and kinematic filters to the Gaia sources. Spatially, candidates must be located within an angular separation $\theta < 5 \times r_{50}$ from the cluster center, where $r_{50}$ is the cluster's half-member radius \citep{Hunt2024}. Astrometrically, candidates were required to lie within a conservatively expanded region in the three-dimensional parallax–proper-motion space occupied by confirmed cluster members. We express this selection schematically as
\begin{equation}
	\left(\pi,\mu_{\alpha},\mu_{\delta}\right)<1.6\times P.
\end{equation}
Here, $P$ collectively represents the full spans of the three astrometric parameters among confirmed cluster members, with each span defined as the maximum minus minimum of the corresponding parameter. The notation $1.6\times P$ denotes the adopted expansion of this member-defined parameter region, and the inequality is shorthand for requiring a candidate's astrometric parameter triplet to fall within the expanded region. The constraints on all three parameters were applied simultaneously. This conservative selection was intended to reduce the risk of excluding potential members owing to observational scatter. To rigorously purify this preliminary sample, we employed the precision membership determination technique used in previous studies \citep[e.g.,][]{OGrady2020, Grondin2024arXiv240704775G, Yan2026}. A detailed description of the method is provided in Appendix~\ref{apa}. Following the selection procedure, we identified a subdwarf star (Gaia DR3 3029973537553031680, hereafter G302) associated with the OC NGC 2437, as shown in Figure~\ref{fig1}. Its spatial coordinates, proper motions, parallax, and radial velocity (RV; see Section~\ref{sec:params} for the spectroscopic analysis) are all consistent with the distributions of the confirmed cluster members.

Previous age estimates for NGC\,2437 span roughly 300--500\,Myr, with individual determinations of 302\,Myr \citep{CantatGaudin2020}, 316\,Myr \citep{Tarricq2021}, 398\,Myr \citep{Jadhav2024}, 431\,Myr \citep{Hunt2024}, 446\,Myr \citep{Angelo2025}, and 501\,Myr \citep{Dias2021}. To obtain a self-consistent estimate, we fitted the cluster CMD with PARSEC~2.0 isochrones \citep{Bressan2012,Nguyen2022} using the $\chi^2$ method of \citet{Miller2026}, which yields an age of \(380 \pm 60~\mathrm{Myr}\). We adopt this age of 380\,Myr for NGC\,2437 throughout the subsequent analysis.

\section{Photometric and Spectral  Parameters} \label{sec:params}
\begin{figure*}[p]
	\centering
	\includegraphics[
	width=\textwidth,
	height=0.80\textheight,
	keepaspectratio
	]{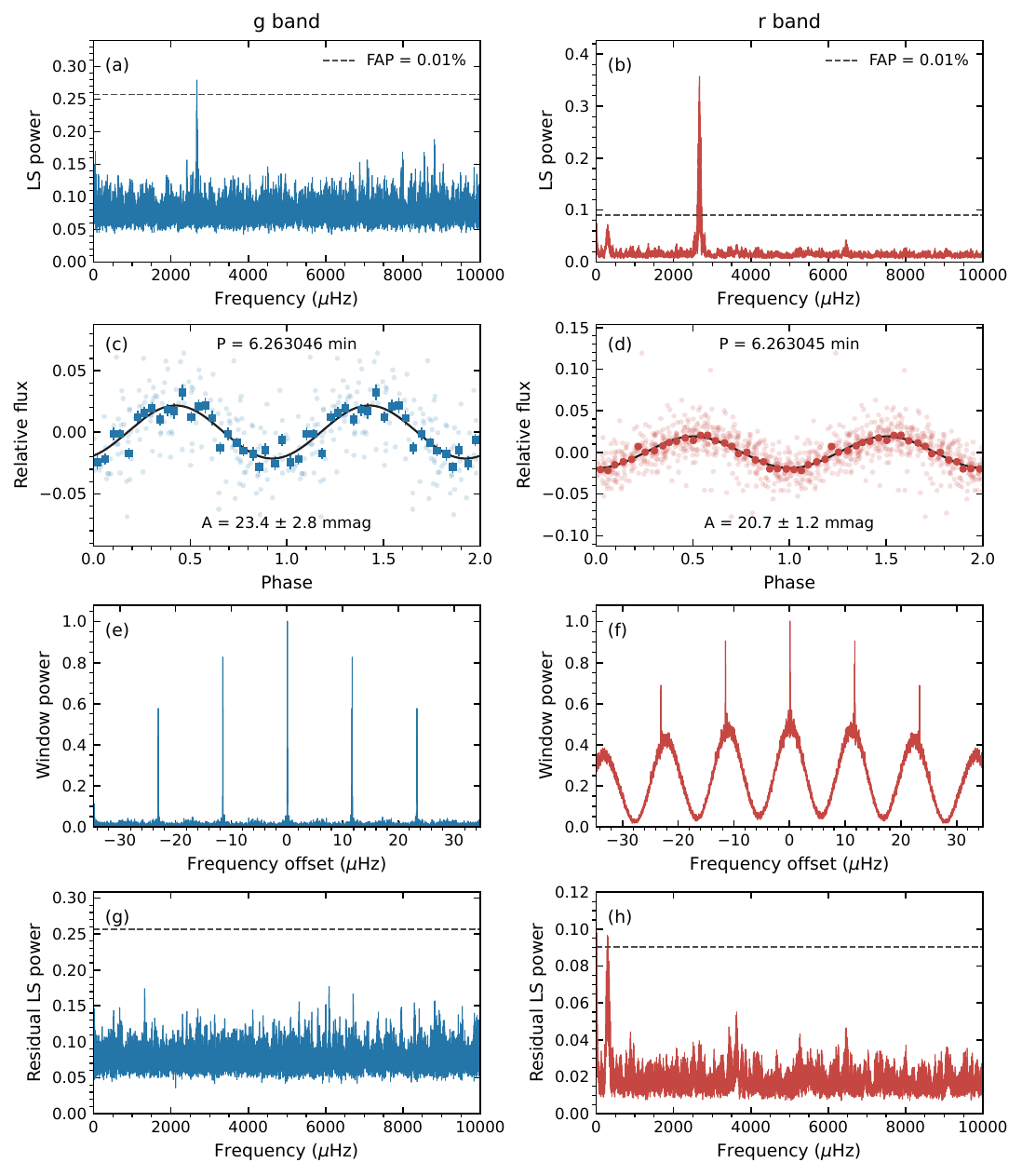}
	\caption{
		ZTF frequency analysis of G302 for the $g$ (left) and $r$ (right)
		bands.
		(a,b) Uncertainty-weighted LS periodograms.
		(c,d) Light curves folded on the dominant period and shown over two
		cycles, plotted as relative flux. Faint points are individual
		measurements, filled symbols are binned weighted means, and black
		curves are the best-fitting sinusoids; $A$ is the semi-amplitude in
		magnitudes. Phase zero points are arbitrary in each band.
		(e,f) Spectral windows, normalized to unity at zero frequency offset.
		(g,h) Residual periodograms after prewhitening the dominant signal.
		Dashed lines in panels (a,b,g,h) mark the 0.01\% FAP level of each
		band.
	}
	\label{fig2}
\end{figure*}

To further characterize G302, we analyzed its Zwicky Transient Facility (ZTF) DR24 light curves \citep{Bellm2019}, using only measurements with \texttt{catflags}~=~0 and valid photometric uncertainties. The final sample contains 176 $g$-band and 538 $r$-band measurements, each spanning about 7.5~yr. The ZTF photometric analysis is presented in Figure~\ref{fig2}.

We computed uncertainty-weighted Lomb--Scargle (LS) periodograms \citep{VanderPlas2018} for the two bands separately over 1.16--10\,000\,$\mu$Hz and evaluated the 0.01\% false-alarm
probability (FAP) level for each band following \citet{Baluev2008}.
Both periodograms are dominated by a peak at $2661.11\,\mu{\rm Hz}$ ($P=6.263$\,min; Figure~\ref{fig2}a,b), with FAPs of $3.5\times10^{-6}$ in $g$ and $2.0\times10^{-45}$ in $r$.
Sinusoidal fits give semi-amplitudes of $23.4\pm2.8$\,mmag in $g$
and $20.7\pm1.2$\,mmag in $r$. The phase-folded light curves show a clear sinusoidal modulation
(Figure~\ref{fig2}c,d), consistent with the short-period,
large-amplitude pulsations hot subdwarfs and related compact pulsators
\citep[e.g.,][]{Kupfer2019,Kupfer2021,Baran2024,Uzundag2024,Krzesinski2025}.

The spectral windows show strong daily aliases (Figure~\ref{fig2}e,f), which account for the side peaks around the
main signal. Because the dominant frequency is recovered
independently in both bands, we identify it as the true pulsation
frequency rather than an alias. After prewhitening the dominant signal, we recomputed the periodograms (Figure~\ref{fig2}g,h).
No significant residual peak remains in the $g$ band.
In the $r$ band, a peak near $1\,{\rm d}^{-1}$ arises from seasonal
photometric offsets, and a weaker peak near 57\,min is not confirmed in the $g$ band and depends on a single densely sampled night.

In addition, we made use of the publicly available archival 
spectroscopy for this target, originally obtained with the Gemini 
Multi-Object Spectrograph on the Gemini South telescope (GMOS-S; 
\citealt{Hook2004}) under program GS-PLAN20210105 (PI: T.\ Kupfer). 
The observations were executed in long-slit mode, utilizing the 
B1200 grating combined with a 0.75~arcsec slit. To improve the signal-to-noise ratio, every five consecutive
exposures were median-combined, yielding 28 spectra that span the	full $\sim$2.4\,h observing sequence. The atmospheric parameters of G302 were determined using the online 
spectroscopic fitting service \textsc{XTgrid} 
\citep{Peter2012}\footnote{\url{http://astroserver.org/}}.  The RV of each combined spectrum was determined with \textsc{XTgrid} as a free parameter, simultaneously with the atmospheric parameters (Table~\ref{tab:obs}). The RV uncertainties were derived in the same way as those of the atmospheric parameters, from the global $\chi^2$ minimization over the optical range, in which \textsc{XTgrid} adjusts the RV shift simultaneously with the atmospheric parameters. Nonlinearities in the wavelength solution therefore propagate into the RV uncertainties and inflate them. The 28 RVs have a mean of $56.3\pm1.5$\,km\,s$^{-1}$, a standard deviation of 8.0\,km\,s$^{-1}$, and a peak-to-peak range of 32.5\,km\,s$^{-1}$. Their scatter exceeds that expected from the formal uncertainties ($\chi^2=97.7$ for 27 degrees of freedom relative to a constant RV), but no independent RV zero-point check was available, so instrumental systematics cannot be excluded. This excess scatter most likely reflects systematics in the wavelength calibration and in the combination of the exposures into 28 bins, rather than intrinsic variations of the star. Consequently, these spectra do not allow us to reliably assess RV variations of G302, whether from pulsation or from orbital motion. The only reliable velocity measure is the mean RV and its standard deviation, which we use for the membership analysis. In addition, because each combined spectrum covers about 4.8\,min, comparable to the 6.263-min pulsation period, any pulsational RV signal is largely averaged out. \citet{Bradshaw2026} measured RVs from the individual spectra and detected a significant signal at the photometric period. Fitting the RV curve phase-folded and binned on this period, they obtained a peak-to-peak pulsational RV amplitude of $18\pm3$\,km\,s$^{-1}$.

\citet{Bradshaw2026} recently conducted a systematic spectroscopic follow-up of hot subdwarf variables discovered by ZTF, and G302 is included in their sample (designated sdBV6). They analyzed the same Gemini/GMOS-S spectra using a hybrid LTE/NLTE model grid. Our independent fit with \textsc{XTgrid} yields atmospheric parameters consistent with theirs within the uncertainties. Table~\ref{tab:G302_compare} compares our best-fit values with those of \citet{Bradshaw2026}, and the Gemini spectral fits are presented in Appendix~\ref{apb}.

\begin{table}[h]
	\centering
	\caption{Comparison of the atmospheric parameters and
		radial velocities of G302 derived in this work with
		those reported by \citet{Bradshaw2026}.}
	\label{tab:G302_compare}
	\begin{tabular}{ccc}
		\hline\hline
		Parameter & This work & \citet{Bradshaw2026} \\
		\hline
		$T_{\rm eff}$ (K) & $28\,317 \pm 708$ & $29\,115 \pm 602$ \\
		$\log g$ (dex) & $5.33 \pm 0.10$ & $5.35 \pm 0.07$ \\
		$\langle \mathrm{RV} \rangle$ (km\,s$^{-1}$)$^a$ & $56.3 \pm 1.5$ & --- \\
		$\Delta \mathrm{RV}$ (km\,s$^{-1}$)$^b$ & --- & $18 \pm 3$ \\
		\hline
	\end{tabular}
	\\[1mm]
	\begin{flushleft}
		$^a$ Mean RV of the 28 combined spectra; the uncertainty is the
		standard error of the mean.\\
		$^b$ Peak-to-peak pulsational RV amplitude.
	\end{flushleft}
\end{table}

\section{The potential Evolution Channel of G302} \label{sec:evolution}

\subsection{Constraints from the cluster age} \label{sec:channel}

As a member of NGC~2437, G302 is expected to have evolved from the zero-age main sequence (ZAMS) to its present stage within $\sim 380$\,Myr. This timescale is difficult to reconcile with the canonical pathway, in which a low-mass star loses its envelope near the tip of the red giant branch (RGB), since such progenitors take several Gyr to reach that stage. The progenitor of G302 is therefore likely an intermediate-mass star. 

The cluster age also provides an indirect constraint on the formation channel. In the BPS models of \citet{Han2002,Han2003}, sdBs form through stable RLOF, CE ejection, or the merger of two helium WDs, with relative contributions that depend strongly on the population age \citep{Han2008}. In populations younger than $\sim 1$\,Gyr, the first stable RLOF channel accounts for $\gtrsim 90\%$ of sdBs, while the CE ejection and merger channels become significant only at $t \gtrsim 1$--2\,Gyr. The first CE ejection channel is also expected to be inefficient for intermediate-mass progenitors. A primary of $\gtrsim 2\,M_\odot$ fills its Roche lobe while its envelope is still massive and tightly bound, so the orbital energy released during the spiral-in may be insufficient to eject the envelope, and the system is likely to merge \citep{Han2008}. The available spectra provide only limited constraints on the binary nature of G302. The short baseline cannot constrain orbital motion on longer timescales, including that expected for post-CE binaries. The infrared excess reported by \citet{Bradshaw2026} indicates a cool late-type companion ($T_{\rm eff} = 3900^{+1500}_{-500}$\,K). Such companions are found in both stable-RLOF and post-CE sdB binaries. No significant reflection effect or eclipse is detected in the residual ZTF periodograms (Section~\ref{sec:params}), which disfavors the shortest-period post-CE configurations with a cool companion, within the ZTF detection limits.

Based mainly on the cluster age and the BPS predictions, we regard the first stable RLOF channel as the most plausible origin of G302, with the mass transfer probably starting while the progenitor was crossing the Hertzsprung gap (HG).
For RGB donors, stable mass transfer typically requires an initial mass ratio of $q \lesssim 1.2$--1.5 \citep{Han2002}. For a $\sim 3\,M_\odot$ progenitor, this would imply a companion of $\gtrsim 2\,M_\odot$, which would likely dominate the optical flux and has not been observed. In contrast, HG donors with radiative envelopes can sustain stable mass transfer at considerably larger mass ratios \citep[$q_{\rm crit} \approx 3$--4;][]{Han2003,Clausen2012,Ge2015}, although a low-mass companion would place the system close to this stability limit.

\subsection{MESA modeling of the progenitor} \label{sec:mesa}

To determine the progenitor in more detail, we computed stellar evolution models with the Modules for Experiments in Stellar Astrophysics (\texttt{MESA}, version r26.4.1; \citealt{Paxton2011,Paxton2013,Paxton2015,Paxton2018,Paxton2019,Jermyn2023}). Because the orbital parameters of G302 are not yet known, we did not model the binary mass transfer directly. Instead, we mimic the stripping with an enhanced stellar wind applied during the HG. This approach reproduces the main outcome of stable RLOF, namely a helium core covered by a thin hydrogen-rich envelope. We computed tracks with initial masses of $2$--$4\,M_\odot$ in steps of $0.1\,M_\odot$.
The enhanced wind is switched on when the star, after leaving the main sequence, cools below a threshold effective temperature within the HG ($\log T_{\rm eff} = 3.70$--3.95). It removes mass at a rate of $10^{-6}$--$10^{-5}\,M_\odot\,{\rm yr}^{-1}$ and is switched off when the remaining hydrogen-rich envelope mass drops to approximately $0.02\,M_\odot$.

\begin{figure*}[ht!] 
	\centering
	\includegraphics[width=\textwidth]{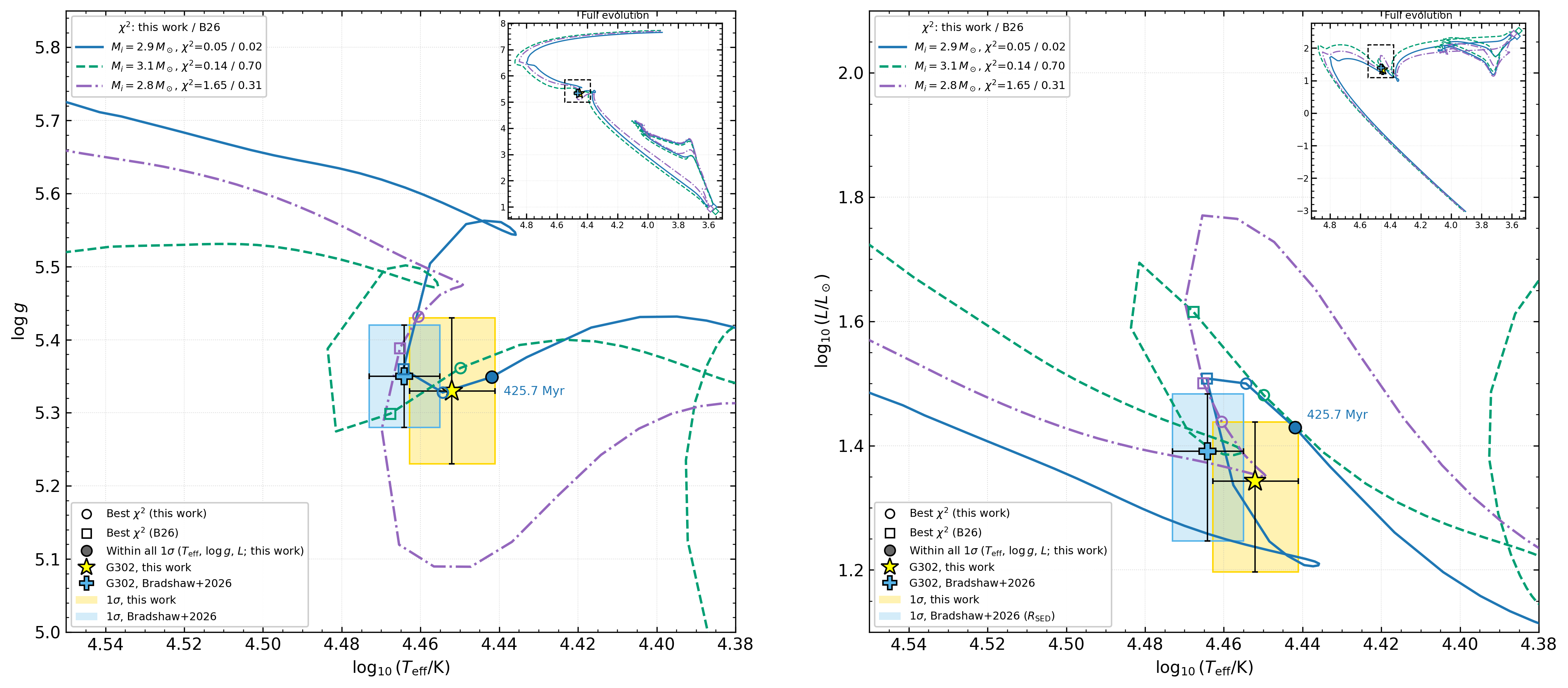}
	\caption{MESA evolutionary tracks compared with the atmospheric parameters of G302 in the Kiel diagram (left) and the HR diagram (right). The yellow star and yellow shaded box show the parameters derived in this work and their marginal \(1\sigma\) ranges. The blue cross and blue shaded box show those of \cite{Bradshaw2026}. The literature luminosity is not given and is estimated from SED radius (\(R=0.195^{+0.020}_{-0.029}\,R_\odot\)) combined with their \(T_{\rm eff}\). Blue solid, green dashed, and purple dash-dotted curves show the tracks with \(M_i=2.9\), \(3.1\), and \(2.8\,M_\odot\), respectively. These are the three best-matching tracks among the  evolution tracks. The legend gives \(\chi^2_{\rm spec}\) for this work and for \cite{Bradshaw2026}. Main panels show only the evolution after the cool turning point, and the insets show the full evolution. Open circles and open squares mark the native models with the lowest \(\chi^2_{\rm spec}\) relative to this work and to \cite{Bradshaw2026}, respectively. In the Kiel diagram, the best-fitting models of the \(2.9\) and \(3.1\,M_\odot\) tracks lie within the \(1\sigma\) ranges of both parameter sets. The best-fitting model of the \(2.8\,M_\odot\) track lies only within the \(1\sigma\) range of \cite{Bradshaw2026}. The filled circle marks the only native model that satisfies the \(1\sigma\) ranges of \(T_{\rm eff}\), \(\log g\), and \(L\) derived in this work simultaneously. It lies on the \(2.9\,M_\odot\) track at \(t=425.7\) Myr, with \(T_{\rm eff}=27\,665\) K, \(\log g=5.35\), \(\log(L/L_\odot)=1.43\), and \(\chi^2_{\rm spec}=0.90\). None of the tracks satisfies all three \(1\sigma\) ranges of \cite{Bradshaw2026} at once.  \label{fig_mesa}}
\end{figure*}

We selected the best-fitting models in two steps. First, we required each track to reach the sdB phase ($\log T_{\rm eff} > 4.35$ and $\log g > 5.0$) within an age window of 300--500\,Myr, which fully covers the cluster age of $380 \pm 60$\,Myr. Only the evolution after the cool turning point of the stripped star, the lowest $T_{\rm eff}$ reached before it moves to the hot side of the HR diagram, was considered. Second, for each of these tracks, we computed at every native model point of the sdB phase within the age window
\begin{equation}
	\chi^2_{\rm spec} = \sum_{i} \left(\frac{x_i - x_{i,{\rm obs}}}{\sigma_{i}}\right)^2 ,
\end{equation}
where $x_i \in \{\log T_{\rm eff},\, \log g\}$, and adopted the point with the minimum $\chi^2_{\rm spec}$. No interpolation between model points or age term was applied. The fit was performed separately for the atmospheric parameters derived in this work ($T_{\rm eff} = 28\,317 \pm 708$\,K, $\log g = 5.33 \pm 0.10$) and those of \citet{Bradshaw2026} ($T_{\rm eff} = 29\,115 \pm 602$\,K, $\log g = 5.35 \pm 0.07$). For each best-fitting point, we also checked whether the model luminosity lies within the $1\sigma$ range of the SED luminosity ($\log L/L_\odot = 1.34^{+0.10}_{-0.15}$). Eleven tracks reach the sdB phase within the age window. The $2.9\,M_\odot$ track gives the best fit for both parameter sets ($\chi^2_{\rm spec,min} = 0.05$ for this work and $0.02$ for \citealt{Bradshaw2026}). The $3.1\,M_\odot$ ($\chi^2_{\rm spec,min} = 0.14$ and 0.70) and $2.8\,M_\odot$ (1.65 and 0.31) tracks also fit well. The remaining tracks are clearly worse, with $\chi^2_{\rm spec,min} \geq 3.0$ for this work and $\geq 4.7$ for \citet{Bradshaw2026}, except the $2.7\,M_\odot$ track, which gives $1.04$ for this work but $4.73$ for \citet{Bradshaw2026}. The progenitor mass of G302 is therefore estimated to be $\approx 2.8$--$3.1\,M_\odot$, with the best fit at $2.9\,M_\odot$. Figure~\ref{fig_mesa} shows these three tracks in the Kiel and HR diagrams.

In the best-fitting model, the $2.9\,M_\odot$ progenitor leaves the main sequence at $t \approx 289$\,Myr. About 11\,Myr later, while crossing the HG, it is stripped from $2.90$ to $\approx 0.41\,M_\odot$ within $\sim 2.3$\,Myr, leaving a $\approx 0.37\,M_\odot$ helium core. Because the core is non-degenerate, helium ignites quietly at $t \approx 303$\,Myr without a helium flash, and the star reaches the EHB at $t \approx 310$\,Myr. At $t \approx 427$\,Myr, consistent with the cluster age within $1\sigma$, the model provides the closest match to G302, with $T_{\rm eff} = 28\,469$\,K, $\log g = 5.33$, $\log L/L_\odot = 1.50$, and $R = 0.231\,R_\odot$ for our parameters. Similarly, it gives $T_{\rm eff} = 29\,132$\,K and $\log g = 5.36$ for those of \citet{Bradshaw2026}. At this point, the sdB has a mass of $0.41\,M_\odot$. Its central helium abundance ($Y_{\rm c} \approx 0.003$) indicates that it is at the very end of core helium burning, which ends at $t \approx 428$\,Myr.

The match near core helium exhaustion is common to our
best-fitting tracks. During most of core helium burning, these models have $T_{\rm eff}\approx22\,000$--24\,000\,K, cooler than G302. They reach the observed $T_{\rm eff}$ and $\log g$ only when the core contracts as helium is exhausted ($Y_{\rm c}\lesssim0.01$), a phase that lasts only $\sim$2--3\,Myr,
or $\sim$2\% of the core-helium-burning lifetime ($\sim$94--124\,Myr). The $3.1$ and $2.8\,M_\odot$ tracks also come closest to the observations at this stage, with core helium exhausted at 349 and 446\,Myr, respectively. Within our models, G302 is therefore in a short, late stage of core helium burning. In this interpretation, the progenitor mass is set mainly by the requirement that core helium exhaustion occurs close to the cluster age, which gives 2.8--3.1\,$M_\odot$. Because the $2.9$ and $3.1\,M_\odot$ tracks both give $\chi^2_{\rm spec}<1$ for both parameter sets, the best-fitting mass of $2.9\,M_\odot$ is not strongly preferred within this range. The corresponding sdB masses are 0.41--0.45\,$M_\odot$. Alternatively, a thinner residual hydrogen envelope would make the sdB hotter during the main core-helium-burning phase, allowing a match at an earlier and longer-lived stage, which could shift the inferred progenitor mass. Asteroseismology could in principle distinguish between these possibilities, but its potential for G302 is currently limited. Constraining the core-helium-burning status or the thickness of the residual hydrogen envelope requires detailed seismic modeling based on the identification of multiple independent pulsation modes (typically $\gtrsim 10$), which is not possible with the single frequency detected so far. Intensive photometric follow-up is therefore needed to search for additional modes and to establish whether G302 is a predominantly p-mode or a hybrid p/g-mode pulsator. While p modes mainly probe the outer layers, the detection of g modes would also constrain the deep interior and the core-helium-burning status, so the actual asteroseismic potential of G302 depends strongly on whether additional modes are detected.

The best-fitting model reproduces the spectroscopic $T_{\rm eff}$ and $\log g$ of both this work and \citet{Bradshaw2026} within $1\sigma$, as well as the cluster age. The remaining tension lies in the luminosity and radius. The model luminosity ($\log L/L_\odot = 1.50$) and radius ($0.231\,R_\odot$) exceed the values derived independently from the SED fit ($\log L/L_\odot = 1.34^{+0.10}_{-0.15}$ and $R = 0.195^{+0.020}_{-0.029}\,R_\odot$) by $\sim 1.6\sigma$ and $\sim 1.8\sigma$, respectively. Because the star evolves rapidly at the end of core helium burning, this tension is sensitive to the exact evolutionary stage. The model $\sim 1$\,Myr earlier ($t = 425.7$\,Myr; filled circle in Figure~\ref{fig_mesa}) has $T_{\rm eff} = 27\,665$\,K, $\log g = 5.35$, and $\log L/L_\odot = 1.43$. It lies within the $1\sigma$ ranges of all three quantities derived in this work, with $\chi^2_{\rm spec} = 0.90$. The model mass of $0.41\,M_\odot$ is higher than the spectroscopic mass of \citet{Bradshaw2026} ($0.31^{+0.08}_{-0.09}\,M_\odot$). Because that mass was derived from $\log g$ and the SED radius, and the model $\log g$ agrees with theirs, the difference mainly reflects the larger model radius. A possible flux contribution from the cool companion and the uncertainty in the distance could both affect the SED radius, and thus the luminosity and spectroscopic mass derived from it.

The use of a single-star model rests on the Vogt--Russell theorem \citep{Kippenhahn2013}, which implies that the structure of a star in hydrostatic and thermal equilibrium, and hence its subsequent evolution, is essentially determined by its mass and internal chemical composition profile, so a single-star model can represent the binary product if it reproduces both. Because the thermal timescale of the stripped star is much shorter than its core-helium-burning lifetime, the remnant quickly relaxes to thermal equilibrium after the envelope is removed, but this does not erase the chemical and structural signatures of the preceding evolution, which are set by the evolutionary stage at which stripping occurs. In our models, the progenitor is evolved self-consistently from the ZAMS into the HG before the stripping is switched on, at the stage where first stable RLOF from an HG donor is expected. The helium-core structure, chemical profile, and mixing history established before stripping are therefore computed rather than imposed. Because both the enhanced wind and RLOF remove mass from the surface, the remnant largely inherits this profile, modified by nuclear burning during the stripping. The equivalence with a binary product therefore extends beyond the global core and envelope masses, but it is not exact. The post-stripping profile depends on the onset, rate, and termination of the mass removal, which are prescribed in our models but set by the mass-transfer history in a binary. Our models also neglect mixing induced by the binary interaction and have not been compared directly with detailed binary evolution calculations. Because the helium core is still growing during the HG, differences in the stripping history can change the helium-core mass and the residual hydrogen envelope, and hence the core-helium-burning lifetime and the inferred progenitor mass. In our grid, tracks with the same initial mass but different stripping parameters indeed give substantially different fits. The range of 2.8--3.1\,$M_\odot$ should therefore be regarded as indicative and conditional on the adopted stripping prescription. The cluster age provides a more robust constraint, restricting the progenitor to intermediate masses, but this constraint is not entirely independent of the stripping. In a binary, the onset of stripping is set by the orbital separation, which our range of threshold temperatures only partly samples. The lower limit of this intermediate-mass range, set mainly by the main-sequence lifetime, is largely insensitive to the stripping, whereas its upper limit, set by the end of core helium burning, depends on the core mass left by the stripping.

The current constraints come mainly from the atmospheric parameters and the cluster age, because the binary parameters of G302 are still unknown. An orbital period, obtained from long-term RV monitoring, would make a complete reconstruction of its evolution possible. Combined with the companion mass from the mass function and a well-constrained SED fit, it would allow the first stable RLOF phase to be modeled with detailed binary evolution calculations instead of an enhanced wind. Such calculations would determine the initial orbital period, the initial mass ratio, and the mass-transfer efficiency. They would also test whether mass transfer from an HG donor remains stable at the large mass ratio implied by a low-mass companion \citep{Han2003,Clausen2012,Ge2015}. Higher-resolution spectroscopy would further reduce the systematic uncertainty in $\log g$. With these data, G302 could serve as an age-calibrated benchmark for binary evolution models of hot subdwarfs.

\section{Conclusion} \label{sec:summary}

We report the first spectroscopically confirmed sdB in a young OC. Using Gaia DR3 astrometry and RVs, we find that the sdB star G302 is a high-confidence member of NGC~2437, whose age we determine to be $\sim 380$~Myr from isochrone fitting. All previously known cluster sdBs reside in systems older than $\sim 6$~Gyr. G302 therefore extends this sample to young populations and has an age that is independent of stellar models of the sdB itself.

The cluster age allows us to estimate the
progenitor mass of an sdB descended from an intermediate-mass star. A low-mass star cannot reach the sdB stage within $\sim 380$~Myr, so
the progenitor should be an intermediate-mass star. By fitting
\texttt{MESA} tracks to the atmospheric parameters of G302 within the cluster age window, we find an initial mass of $\approx 2.8$--$3.1\,M_\odot$ for the adopted stripping prescription.
In these models, the star is stripped in the HG, ignites helium
non-degenerately, and becomes a 0.41--0.45\,$M_\odot$ sdB. The models reproduce the observed atmospheric parameters only near the end of core helium burning, so the inferred mass mainly reflects the requirement that this stage coincides with the cluster age. The best-fitting $2.9\,M_\odot$ model reproduces the spectroscopic $T_{\rm eff}$ and $\log g$ of both studies within $1\sigma$. Its luminosity and radius are slightly higher than the SED values, by $\lesssim 2\sigma$.

The cluster age also informs the likely formation channel. At this age,  published BPS predicts that most sdBs form through the first stable RLOF channel. The current spectra do not constrain the orbit, and the infrared excess from a late-type companion is consistent with, but does not uniquely require, this channel. We therefore consider first stable RLOF, with mass transfer beginning in the HG, to be the most likely formation channel, while noting that this conclusion relies mainly on the cluster age and model predictions.

G302 is thus an sdB whose age and progenitor mass are constrained and whose formation channel can be directly tested, making it a valuable benchmark for hot subdwarf evolution. Long-term RV monitoring, higher-resolution spectroscopy, and asteroseismology will further test this scenario.

\begin{acknowledgments}
	C.Y.L. and H.H.Y. acknowledge support from the National Natural Science Foundation of China (NSFC) through grant 12233013. X.M. acknowledges support from 
	National Natural Science Foundation of China (Nos. 
	12288102, 12333008), the Strategic Priority Research 
	Program of the Chinese Academy of Sciences (Grant Nos. 
	XDB1160303, XDB1160000), and the National Science 
	Foundation of China and National Key R\&D Program of 
	China (No. 2021YFA1600403), Yunnan Fundamental 
	Research Projects (No. 202401BC070007), International 
	Centre of Supernovae, Yunnan Key Laboratory (No. 
	202302AN360001), Yunnan Revitalization Talent Support 
	Program – Yunling Scholar Project and the Yunnan 
	Revitalization Talent Support Program – Science \& 
	Technology Champion Project (No. 202305AB350003), 
	and the China Manned Space Program with
	grant No. CMS-CSST-2025-A13.
	D.R.M acknowledges support from the Natural Sciences and 
	Engineering Research Council of Canada Discovery Grants 
	DG-RGPIN-2022-03051 and DG-RGPIN-2023-04486. H.H.Y. thanks the China Association for Science and Technology (CAST) Young Elite Scientists Sponsorship Program for support. J.K.Z acknowledges support from the NSFC (Grant No. 12273055).  This research has used the services of \href{https://astroserver.org}{www.Astroserver.org} under reference A95MVJ. The MESA  files used in this work
	are publicly available on Zenodo at
	\dataset[doi:10.5281/zenodo.23859099.]{https://zenodo.org/records/23859100}

\end{acknowledgments}

\appendix

\section{Appendix A} \label{apa}

The method for further refining cluster membership is as follows. We modeled the intrinsic distribution of each cluster by constructing two independent covariance matrices ($C$). For the 2D kinematic analysis, the parameter vector $\boldsymbol{\mu}$ and its corresponding covariance matrix $C$ are defined as:
\begin{equation}
	\boldsymbol{\mu} = [\mu_\alpha, \mu_\delta ]
\end{equation}
\begin{equation}
	C = \begin{pmatrix}
		c_{\mu_\alpha\mu_\alpha} & c_{\mu_\alpha\mu_\delta} \\ 
		c_{\mu_\delta\mu_\alpha} & c_{\mu_\delta\mu_\delta}  \\ 
	\end{pmatrix}.
\end{equation}
Similarly, for the 3D spatial analysis, the vector includes equatorial coordinates and parallax:
\begin{equation}
	\boldsymbol{\mu} = [\alpha, \delta, \pi]
\end{equation}
with its covariance matrix defined as:
\begin{equation}
	C = \begin{pmatrix}
		c_{\alpha\alpha} & c_{\alpha\delta} & c_{\alpha\pi} \\ 
		c_{\delta\alpha} & c_{\delta\delta} & c_{\delta\pi}  \\ 
		c_{\pi\alpha} & c_{\pi\delta} & c_{\pi\pi} 
	\end{pmatrix}.
\end{equation}

To deconvolve the intrinsic cluster spread from individual measurement uncertainties, we modeled the observed parameter vector of each cluster member as drawn from a Gaussian whose covariance is the sum of the intrinsic cluster covariance and the measurement covariance of that star,
\begin{equation}
	C_{{\rm tot},i} = C + C_{{\rm err},i},
\end{equation}
where $C$ is the intrinsic covariance matrix of the cluster defined above, and $C_{{\rm err},i}$ is the measurement covariance matrix of the $i$-th star, constructed from its Gaia DR3 uncertainties and correlation coefficients. The optimal intrinsic covariance is then determined by maximizing the log-likelihood
\begin{equation}
	\ln L = -\frac{1}{2} \sum_{i} \left[ \left( \boldsymbol{\mu}_i - \boldsymbol{\mu} \right)^{T} C_{{\rm tot},i}^{-1} \left( \boldsymbol{\mu}_i - \boldsymbol{\mu} \right) + \ln \det\!\left( 2\pi C_{{\rm tot},i} \right) \right],
\end{equation}
where $\boldsymbol{\mu}_i$ is the measurement vector of the $i$-th member and $\boldsymbol{\mu}$ is the cluster mean. We used \texttt{scipy.optimize.minimize} to maximize $\ln L$ and obtain the optimized intrinsic covariance matrix, denoted as $\boldsymbol{C_{*}}$, for both the 2D kinematic and 3D spatial domains.

Finally, we evaluated the membership quality of each hot subdwarf candidate using the $\chi^2$ statistic, which measures its distance from the cluster median $\boldsymbol{X}$ in the covariance-weighted parameter space:
\begin{equation}
	\chi^2 = (\boldsymbol{\mu} - \boldsymbol{X})^T \boldsymbol{C}_{*}^{-1}(\boldsymbol{\mu} - \boldsymbol{X}).
\end{equation}

We applied threshold criteria corresponding to a 99.5\% confidence interval: candidates are considered members if they satisfy $\chi^2 < 12.8$ in the 3D spatial domain and $\chi^2 < 10.6$ in the 2D kinematic domain. Furthermore, candidates that simultaneously satisfy both the 2D kinematic and 3D spatial criteria are defined as our high-confidence 5D member sample.

\section{Appendix B}\label{apb}
\begin{figure*}[ht!]
	\centering
	\includegraphics[width=0.85\textwidth]{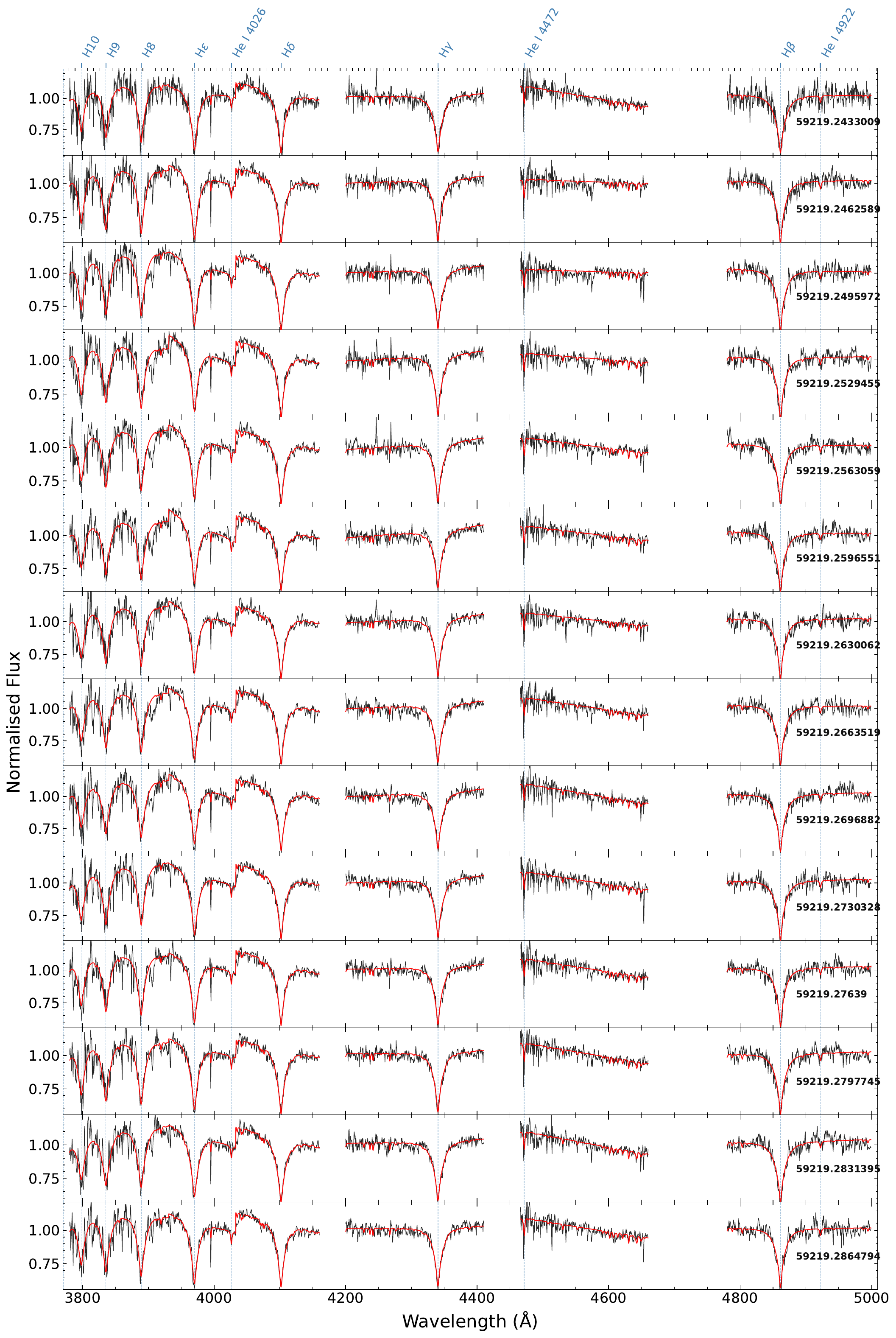}
\caption{The observed spectrum of the target star (grey line) overlaid with the best-fit synthetic model spectrum (red solid line).}
	\label{spec1}
\end{figure*}

\begin{figure*}[ht!]
	\centering
	\includegraphics[width=0.85\textwidth]{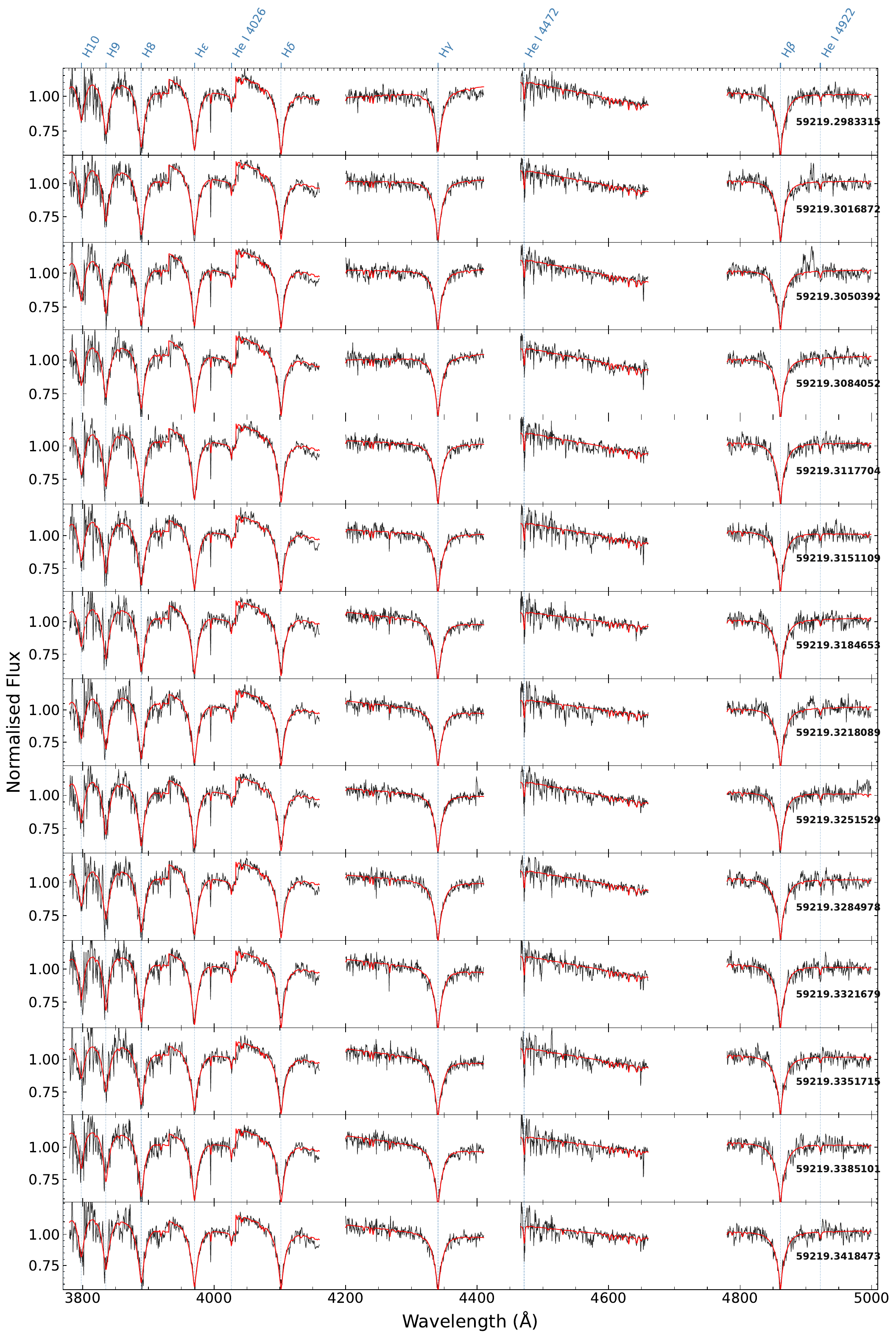}
	\caption{Same as Figure \ref{spec1}.}
	\label{spec2}
\end{figure*}

This appendix presents the spectral fitting (Figures~\ref{spec1}
and~\ref{spec2}) and the fitting parameters listed in
Table~\ref{tab:obs}. For the spectral analysis, we fitted each of the 28 combined GMOS-S spectra independently with \textsc{XTgrid} \citep{Peter2012}. \textsc{XTgrid} is a steepest-descent $\chi^{2}$ 
minimization procedure built around the non-LTE model atmosphere and 
spectral synthesis codes \textsc{Tlusty} and \textsc{Synspec} 
\citep{Hubeny1995, Lanz2003}. Starting from an initial model derived 
from a rapid spectral-type estimate, the procedure computes synthetic 
spectra on-the-fly and iteratively updates all free parameters, namely 
$T_{\rm eff}$, $\log g$, the helium abundance 
$\log(N_{\rm He}/N_{\rm H})$, the projected rotational velocity 
$v\sin i$, and the RV, along the steepest gradient of the global $\chi^{2}$ 
surface until convergence is reached. Because the synthetic spectra 
are generated on demand rather than interpolated from a pre-computed 
grid, the method covers the full parameter space of hot subdwarfs 
without grid-edge artifacts. The statistical uncertainties of the 
best-fit parameters, including the RV, are estimated from the $\chi^{2}$ 
surface around the minimum. Because the RV shift is optimized together with the atmospheric parameters in the global fit, nonlinearities in the wavelength solution propagate into the RV uncertainties. The fits to the individual spectra are shown in Figures~\ref{spec1} and~\ref{spec2}, and the resulting parameters are listed in Table~\ref{tab:obs}. The mean values adopted in Section~\ref{sec:params} are the averages of these 28 fits.

\begin{deluxetable}{lcccccc}
	\tablecaption{Spectroscopic Fitting Parameters of G302 \label{tab:obs}}
	\tablehead{
		\colhead{MJD} & \colhead{$T_{\rm eff}$} & \colhead{$\sigma_{T_{\rm eff}}$} & \colhead{$\log g$} & \colhead{$\sigma_{\log g}$} & \colhead{RV} & \colhead{$\sigma_{\rm RV}$} \\
		\colhead{} & \colhead{(K)} & \colhead{(K)} & \colhead{(dex)} & \colhead{(dex)} & \colhead{(km\,s$^{-1}$)} & \colhead{(km\,s$^{-1}$)}
	}
	\startdata
	59219.2433009 & 27190 & 680 & 5.388 & 0.069 & 49.88 & 3.75 \\
	59219.2462589 & 27440 & 420 & 5.210 & 0.057 & 58.60 & 10.54 \\
	59219.2495972 & 27150 & 250 & 5.257 & 0.090 & 43.79 & 4.68 \\
	59219.2529455 & 28540 & 140 & 5.205 & 0.067 & 43.70 & 4.875 \\
	59219.2563059 & 28720 & 290 & 5.277 & 0.035 & 61.05 & 47.97 \\
	59219.2596551 & 29170 & 250 & 5.337 & 0.057 & 61.29 & 7.025 \\
	59219.2630062 & 28010 & 190 & 5.269 & 0.047 & 46.88 & 7.025 \\
	59219.2663519 & 28110 & 640 & 5.451 & 0.104 & 64.66 & 14.055 \\
	59219.2696882 & 28590 & 770 & 5.361 & 0.078 & 48.55 & 37.47 \\
	59219.2730328 & 27970 & 500 & 5.206 & 0.127 & 44.64 & 4.10 \\
	59219.27639   & 28370 & 210 & 5.210 & 0.017 & 41.13 & 18.74 \\
	59219.2797745 & 27660 & 310 & 5.249 & 0.043 & 64.81 & 4.68 \\
	59219.2831395 & 28800 & 540 & 5.439 & 0.083 & 66.05 & 4.685 \\
	59219.2864794 & 28640 & 400 & 5.189 & 0.107 & 62.43 & 4.68 \\
	59219.2983315 & 29360 & 570 & 5.420 & 0.063 & 63.97 & 18.74 \\
	59219.3016872 & 28630 & 290 & 5.387 & 0.076 & 57.02 & 2.49 \\
	59219.3050392 & 28920 & 280 & 5.288 & 0.038 & 62.44 & 7.025 \\
	59219.3084052 & 29180 & 330 & 5.336 & 0.029 & 54.24 & 4.68 \\
	59219.3117704 & 28180 & 320 & 5.302 & 0.046 & 54.83 & 9.37 \\
	59219.3151109 & 28720 & 400 & 5.318 & 0.044 & 56.88 & 3.51 \\
	59219.3184653 & 28510 & 270 & 5.450 & 0.086 & 64.27 & 4.095 \\
	59219.3218089 & 27780 & 420 & 5.350 & 0.052 & 55.63 & 8.245 \\
	59219.3251529 & 28240 & 430 & 5.159 & 0.052 & 61.27 & 5.70 \\
	59219.3284978 & 28630 & 470 & 5.492 & 0.061 & 55.87 & 4.68 \\
	59219.3321679 & 26660 & 490 & 5.241 & 0.052 & 73.63 & 2.34 \\
	59219.3351715 & 29760 & 110 & 5.549 & 0.027 & 52.93 & 3.22 \\
	59219.3385101 & 28220 & 210 & 5.449 & 0.030 & 52.08 & 3.51 \\
	59219.3418473 & 27720 & 330 & 5.334 & 0.052 & 54.22 & 5.855 \\
	\enddata
\end{deluxetable}


\bibliography{sample701}{}

@ARTICLE{Heber2016,
	author = {{Heber}, U.},
	title = "{Hot Subluminous Stars}",
	journal = {\pasp},
	year = 2016,
	month = aug,
	volume = {128},
	number = {966},
	pages = {082001},
	doi = {10.1088/1538-3873/128/966/082001},
	archivePrefix = {arXiv},
	eprint = {1604.07749},
	primaryClass = {astro-ph.SR},
	adsurl = {https://ui.adsabs.harvard.edu/abs/2016PASP..128h2001H}
}

@ARTICLE{Geier2017,
	author = {{Geier}, S. and {{\O}stensen}, R.~H. and {Nemeth}, P. and {Gentile Fusillo}, N.~P. and {G{\"a}nsicke}, B.~T. and {Telting}, J.~H. and {Green}, E.~M. and {Schaffenroth}, J.},
	title = "{The population of hot subdwarf stars studied with Gaia. I. The catalog of known hot subdwarf stars}",
	journal = {\aap},
	year = 2017,
	month = apr,
	volume = {600},
	eid = {A50},
	pages = {A50},
	doi = {10.1051/0004-6361/201630135},
	archivePrefix = {arXiv},
	eprint = {1612.02995},
	primaryClass = {astro-ph.SR},
	adsurl = {https://ui.adsabs.harvard.edu/abs/2017A&A...600A..50G}
}

@ARTICLE{Maxted2001,
	author = {{Maxted}, P.~F.~L. and {Heber}, U. and {Marsh}, T.~R. and {North}, R.~C.},
	title = "{The binary fraction of extreme horizontal branch stars}",
	journal = {\mnras},
	year = 2001,
	month = oct,
	volume = {326},
	number = {4},
	pages = {1391-1402},
	doi = {10.1111/j.1365-2966.2001.04714.x},
	archivePrefix = {arXiv},
	eprint = {astro-ph/0103342},
	primaryClass = {astro-ph},
	adsurl = {https://ui.adsabs.harvard.edu/abs/2001MNRAS.326.1391M}
}

@ARTICLE{Stark2003,
	author = {{Stark}, M.~A. and {Wade}, Richard A.},
	title = "{Single and Composite Hot Subdwarf Stars in the Light of 2MASS Photometry}",
	journal = {\aj},
	year = 2003,
	month = sep,
	volume = {126},
	number = {3},
	pages = {1455-1471},
	doi = {10.1086/377017},
	archivePrefix = {arXiv},
	eprint = {astro-ph/0305511},
	primaryClass = {astro-ph},
	adsurl = {https://ui.adsabs.harvard.edu/abs/2003AJ....126.1455S}
}

@ARTICLE{He2025MNRAS,
	author = {{He}, Ruijie and {Meng}, Xiangcun and {Lei}, Zhenxin},
	title = "{The fractions of reflection effect systems detected in different subclasses of hot subdwarfs}",
	journal = {\mnras},
	year = 2025,
	month = nov,
	volume = {543},
	number = {3},
	pages = {2243-2254},
	doi = {10.1093/mnras/staf1612},
	archivePrefix = {arXiv},
	eprint = {2509.17056},
	primaryClass = {astro-ph.SR},
	adsurl = {https://ui.adsabs.harvard.edu/abs/2025MNRAS.543.2243H}
}

@ARTICLE{He2025AA,
	author = {{He}, Ruijie and {Meng}, Xiangcun and {Lei}, Zhenxin and {Yan}, Huahui and {Lan}, Shunyi},
	title = "{Radial velocity variability fractions of different types of hot subdwarf stars}",
	journal = {\aap},
	year = 2025,
	month = jan,
	volume = {693},
	eid = {A121},
	pages = {A121},
	doi = {10.1051/0004-6361/202451411},
	archivePrefix = {arXiv},
	eprint = {2412.13963},
	primaryClass = {astro-ph.SR},
	adsurl = {https://ui.adsabs.harvard.edu/abs/2025A&A...693A.121H}
}

@ARTICLE{Zhang2012,
	author = {{Zhang}, Xianfei and {Jeffery}, C. Simon},
	title = "{Evolutionary models for double helium white dwarf mergers and the formation of helium-rich hot subdwarfs}",
	journal = {\mnras},
	year = 2012,
	month = jan,
	volume = {419},
	number = {1},
	pages = {452-464},
	doi = {10.1111/j.1365-2966.2011.19711.x},
	adsurl = {https://ui.adsabs.harvard.edu/abs/2012MNRAS.419..452Z}
}

@ARTICLE{Justham2011,
	author = {{Justham}, Stephen and {Podsiadlowski}, Philipp and {Han}, Zhanwen},
	title = "{On the formation of single and binary helium-rich subdwarf O stars}",
	journal = {\mnras},
	year = 2011,
	month = jan,
	volume = {410},
	number = {2},
	pages = {984-993},
	doi = {10.1111/j.1365-2966.2010.17497.x},
	archivePrefix = {arXiv},
	eprint = {1008.1584},
	primaryClass = {astro-ph.SR},
	adsurl = {https://ui.adsabs.harvard.edu/abs/2011MNRAS.410..984J}
}

@ARTICLE{Byrne2018,
	author = {{Byrne}, Conor M. and {Jeffery}, C. Simon},
	title = "{Post-common envelope binary stars, radiative levitation, and blue large-amplitude pulsators}",
	journal = {\mnras},
	year = 2018,
	month = dec,
	volume = {481},
	number = {3},
	pages = {3810-3820},
	doi = {10.1093/mnras/sty2545},
	archivePrefix = {arXiv},
	eprint = {1809.04183},
	primaryClass = {astro-ph.SR},
	adsurl = {https://ui.adsabs.harvard.edu/abs/2018MNRAS.481.3810B}
}

@ARTICLE{Pietrukowicz2017,
	author = {{Pietrukowicz}, Pawe{\l} and {Dziembowski}, Wojciech A. and {Latour}, Marilyn and {Angeloni}, Rodolfo and {Poleski}, Rados{\l}aw and {di Mille}, Francesco and {Soszy{\'n}ski}, Igor and {Udalski}, Andrzej and {Szyma{\'n}ski}, Micha{\l} K. and {Wyrzykowski}, {\L}ukasz and {Koz{\l}owski}, Szymon and {Skowron}, Jan and {Skowron}, Dorota and {Mr{\'o}z}, Przemek and {Pawlak}, Micha{\l} and {Ulaczyk}, Krzysztof},
	title = "{Blue large-amplitude pulsators as a new class of variable stars}",
	journal = {Nature Astronomy},
	year = 2017,
	month = aug,
	volume = {1},
	eid = {0166},
	pages = {0166},
	doi = {10.1038/s41550-017-0166},
	archivePrefix = {arXiv},
	eprint = {1706.07802},
	primaryClass = {astro-ph.SR},
	adsurl = {https://ui.adsabs.harvard.edu/abs/2017NatAs...1E.166P}
}

@ARTICLE{Sanjayan2022,
	author = {{Sanjayan}, S. and {Baran}, A.~S. and {Ostrowski}, J. and {N{\'e}meth}, P. and {Pelisoli}, I. and {{\O}stensen}, R. and {Kern}, J.~W. and {Reed}, M.~D. and {Sahoo}, S.~K.},
	title = "{Pulsating subdwarf B stars in the oldest open cluster NGC 6791}",
	journal = {\mnras},
	year = 2022,
	month = jan,
	volume = {509},
	number = {1},
	pages = {763-777},
	doi = {10.1093/mnras/stab2985},
	archivePrefix = {arXiv},
	eprint = {2111.09873},
	primaryClass = {astro-ph.SR},
	adsurl = {https://ui.adsabs.harvard.edu/abs/2022MNRAS.509..763S}
}

@ARTICLE{Kaluzny1992,
	author = {{Kaluzny}, J. and {Udalski}, A.},
	title = "{Photometric Study of the Old Open Cluster NGC 6791}",
	journal = {\actaa},
	year = 1992,
	month = jan,
	volume = {42},
	pages = {29-47},
	adsurl = {https://ui.adsabs.harvard.edu/abs/1992AcA....42...29K}
}

@ARTICLE{Liebert1994,
	author = {{Liebert}, James and {Saffer}, Rex A. and {Green}, Elizabeth M.},
	title = "{The Evolved Hot Stars of The Old, Metal-Rich Galactic Cluster NGC 6791}",
	journal = {\aj},
	year = 1994,
	month = apr,
	volume = {107},
	pages = {1408},
	doi = {10.1086/116954},
	adsurl = {https://ui.adsabs.harvard.edu/abs/1994AJ....107.1408L}
}

@MISC{Landsman2003,
	author = {{Landsman}, W.},
	title = "{The Hot Subdwarfs of the Old, Open Cluster NGC 188}",
	howpublished = {FUSE Proposal ID D165},
	year = 2003,
	month = jan,
	adsurl = {https://ui.adsabs.harvard.edu/abs/2003fuse.prop.D165L}
}

@ARTICLE{Rani2021,
	author = {{Rani}, Sharmila and {Subramaniam}, Annapurni and {Pandey}, Sindhu and {Sahu}, Snehalata and {Mondal}, Chayan and {Pandey}, Gajendra},
	title = "{UOCS. V. UV study of the old open cluster NGC 188 using AstroSat}",
	journal = {Journal of Astrophysics and Astronomy},
	year = 2021,
	month = oct,
	volume = {42},
	number = {2},
	eid = {47},
	pages = {47},
	doi = {10.1007/s12036-020-09683-2},
	archivePrefix = {arXiv},
	eprint = {2012.00510},
	primaryClass = {astro-ph.SR},
	adsurl = {https://ui.adsabs.harvard.edu/abs/2021JApA...42...47R}
}

@ARTICLE{Latour2018,
	author = {{Latour}, Marilyn and {Randall}, Suzanna K. and {Calamida}, Annalisa and {Geier}, Stephan and {Moehler}, Sabine},
	title = "{SHOTGLAS. I. The ultimate spectroscopic census of extreme horizontal branch stars in {\ensuremath{\omega}} Centauri}",
	journal = {\aap},
	year = 2018,
	month = oct,
	volume = {618},
	eid = {A15},
	pages = {A15},
	doi = {10.1051/0004-6361/201833129},
	archivePrefix = {arXiv},
	eprint = {1806.09436},
	primaryClass = {astro-ph.SR},
	adsurl = {https://ui.adsabs.harvard.edu/abs/2018A&A...618A..15L}
}

@ARTICLE{Moehler2007,
	author = {{Moehler}, S. and {Dreizler}, S. and {Lanz}, T. and {Bono}, G. and {Sweigart}, A.~V. and {Calamida}, A. and {Monelli}, M. and {Nonino}, M.},
	title = "{The hottest horizontal-branch stars in {\ensuremath{\omega}} Centauri. Late hot flasher vs. helium enrichment}",
	journal = {\aap},
	year = 2007,
	month = nov,
	volume = {475},
	number = {2},
	pages = {L5-L8},
	doi = {10.1051/0004-6361:20078509},
	archivePrefix = {arXiv},
	eprint = {0709.2839},
	primaryClass = {astro-ph},
	adsurl = {https://ui.adsabs.harvard.edu/abs/2007A&A...475L...5M}
}

@ARTICLE{MoniBidin2006,
	author = {{MoniBidin}, C. and {Moehler}, S. and {Piotto}, G. and {Recio-Blanco}, A. and {Momany}, Y. and {M{\'e}ndez}, R.~A.},
	title = "{The lack of close binaries among hot horizontal branch stars in <ASTROBJ>NGC 6752</ASTROBJ>}",
	journal = {\aap},
	year = 2006,
	month = may,
	volume = {451},
	number = {2},
	pages = {499-513},
	doi = {10.1051/0004-6361:20053940},
	archivePrefix = {arXiv},
	eprint = {astro-ph/0602075},
	primaryClass = {astro-ph},
	adsurl = {https://ui.adsabs.harvard.edu/abs/2006A&A...451..499M}
}

@ARTICLE{Nedhath2025,
	author = {{Nedhath}, Sneha and {Rani}, Sharmila and {Subramaniam}, Annapurni and {Pancino}, Elena},
	title = "{AstroSat/UVIT Study of NGC 663: First detection of Be+sdOB systems in a young star cluster}",
	journal = {\aap},
	year = 2025,
	month = jun,
	volume = {699},
	eid = {L1},
	pages = {L1},
	doi = {10.1051/0004-6361/202555495},
	archivePrefix = {arXiv},
	eprint = {2506.08126},
	primaryClass = {astro-ph.SR},
	adsurl = {https://ui.adsabs.harvard.edu/abs/2025A&A...699L...1N}
}

@ARTICLE{GaiaCollaboration2023A&A...674A...1G,
	author = {{Gaia Collaboration} and {Vallenari}, A. and {Brown}, A.~G.~A. and {Prusti}, T. and {de Bruijne}, J.~H.~J. and {Arenou}, F. and {Babusiaux}, C. and {Biermann}, M. and {Creevey}, O.~L. and {Ducourant}, C. and {Evans}, D.~W. and {Eyer}, L. and {Guerra}, R. and {Hutton}, A. and {Jordi}, C. and {Klioner}, S.~A. and {Lammers}, U.~L. and {Lindegren}, L. and {Luri}, X. and {Mignard}, F. and {Panem}, C. and {Pourbaix}, D. and {Randich}, S. and {Sartoretti}, P. and {Soubiran}, C. and {Tanga}, P. and {Walton}, N.~A. and {Bailer-Jones}, C.~A.~L. and {Bastian}, U. and {Drimmel}, R. and {Jansen}, F. and {Katz}, D. and {Lattanzi}, M.~G. and {van Leeuwen}, F. and {Bakker}, J. and {Cacciari}, C. and {Casta{\~n}eda}, J. and {De Angeli}, F. and {Fabricius}, C. and {Fouesneau}, M. and {Fr{\'e}mat}, Y. and {Galluccio}, L. and {Guerrier}, A. and {Heiter}, U. and {Masana}, E. and {Messineo}, R. and {Mowlavi}, N. and {Nicolas}, C. and {Nienartowicz}, K. and {Pailler}, F. and {Panuzzo}, P. and {Riclet}, F. and {Roux}, W. and {Seabroke}, G.~M. and {Sordo}, R. and {Th{\'e}venin}, F. and {Gracia-Abril}, G. and {Portell}, J. and {Teyssier}, D. and {Altmann}, M. and {Andrae}, R. and {Audard}, M. and {Bellas-Velidis}, I. and {Benson}, K. and {Berthier}, J. and {Blomme}, R. and {Burgess}, P.~W. and {Busonero}, D. and {Busso}, G. and {C{\'a}novas}, H. and {Carry}, B. and {Cellino}, A. and {Cheek}, N. and {Clementini}, G. and {Damerdji}, Y. and {Davidson}, M. and {de Teodoro}, P. and {Nu{\~n}ez Campos}, M. and {Delchambre}, L. and {Dell'Oro}, A. and {Esquej}, P. and {Fern{\'a}ndez-Hern{\'a}ndez}, J. and {Fraile}, E. and {Garabato}, D. and {Garc{\'\i}a-Lario}, P. and {Gosset}, E. and {Haigron}, R. and {Halbwachs}, J.-L. and {Hambly}, N.~C. and {Harrison}, D.~L. and {Hern{\'a}ndez}, J. and {Hestroffer}, D. and {Hodgkin}, S.~T. and {Holl}, B. and {Jan{\ss}en}, K. and {Jevardat de Fombelle}, G. and {Jordan}, S. and {Krone-Martins}, A. and {Lanzafame}, A.~C. and {L{\"o}ffler}, W. and {Marchal}, O. and {Marrese}, P.~M. and {Moitinho}, A. and {Muinonen}, K. and {Osborne}, P. and {Pancino}, E. and {Pauwels}, T. and {Recio-Blanco}, A. and {Reyl{\'e}}, C. and {Riello}, M. and {Rimoldini}, L. and {Roegiers}, T. and {Rybizki}, J. and {Sarro}, L.~M. and {Siopis}, C. and {Smith}, M. and {Sozzetti}, A. and {Utrilla}, E. and {van Leeuwen}, M. and {Abbas}, U. and {{\'A}brah{\'a}m}, P. and {Abreu Aramburu}, A. and {Aerts}, C. and {Aguado}, J.~J. and {Ajaj}, M. and {Aldea-Montero}, F. and {Altavilla}, G. and {{\'A}lvarez}, M.~A. and {Alves}, J. and {Anders}, F. and {Anderson}, R.~I. and {Anglada Varela}, E. and {Antoja}, T. and {Baines}, D. and {Baker}, S.~G. and {Balaguer-N{\'u}{\~n}ez}, L. and {Balbinot}, E. and {Balog}, Z. and {Barache}, C. and {Barbato}, D. and {Barros}, M. and {Barstow}, M.~A. and {Bartolom{\'e}}, S. and {Bassilana}, J.-L. and {Bauchet}, N. and {Becciani}, U. and {Bellazzini}, M. and {Berihuete}, A. and {Bernet}, M. and {Bertone}, S. and {Bianchi}, L. and {Binnenfeld}, A. and {Blanco-Cuaresma}, S. and {Blazere}, A. and {Boch}, T. and {Bombrun}, A. and {Bossini}, D. and {Bouquillon}, S. and {Bragaglia}, A. and {Bramante}, L. and {Breedt}, E. and {Bressan}, A. and {Brouillet}, N. and {Brugaletta}, E. and {Bucciarelli}, B. and {Burlacu}, A. and {Butkevich}, A.~G. and {Buzzi}, R. and {Caffau}, E. and {Cancelliere}, R. and {Cantat-Gaudin}, T. and {Carballo}, R. and {Carlucci}, T. and {Carnerero}, M.~I. and {Carrasco}, J.~M. and {Casamiquela}, L. and {Castellani}, M. and {Castro-Ginard}, A. and {Chaoul}, L. and {Charlot}, P. and {Chemin}, L. and {Chiaramida}, V. and {Chiavassa}, A. and {Chornay}, N. and {Comoretto}, G. and {Contursi}, G. and {Cooper}, W.~J. and {Cornez}, T. and {Cowell}, S. and {Crifo}, F. and {Cropper}, M. and {Crosta}, M. and {Crowley}, C. and {Dafonte}, C. and {Dapergolas}, A. and {David}, M. and {David}, P. and {de Laverny}, P. and {De Luise}, F. and {De March}, R.},
	title = "{Gaia Data Release 3. Summary of the content and survey properties}",
	journal = {\aap},
	year = 2023,
	month = jun,
	volume = {674},
	eid = {A1},
	pages = {A1},
	doi = {10.1051/0004-6361/202243940},
	archivePrefix = {arXiv},
	eprint = {2208.00211},
	primaryClass = {astro-ph.GA},
	adsurl = {https://ui.adsabs.harvard.edu/abs/2023A&A...674A...1G}
}

@ARTICLE{Hunt2023,
	author = {{Hunt}, Emily L. and {Reffert}, Sabine},
	title = "{Improving the open cluster census. II. An all-sky cluster catalogue with Gaia DR3}",
	journal = {\aap},
	year = 2023,
	month = may,
	volume = {673},
	eid = {A114},
	pages = {A114},
	doi = {10.1051/0004-6361/202346285},
	archivePrefix = {arXiv},
	eprint = {2303.13424},
	primaryClass = {astro-ph.GA},
	adsurl = {https://ui.adsabs.harvard.edu/abs/2023A&A...673A.114H}
}

@ARTICLE{Hunt2024,
	author = {{Hunt}, Emily L. and {Reffert}, Sabine},
	title = "{Improving the open cluster census. III. Using cluster masses, radii, and dynamics to create a cleaned open cluster catalogue}",
	journal = {\aap},
	year = 2024,
	month = jun,
	volume = {686},
	eid = {A42},
	pages = {A42},
	doi = {10.1051/0004-6361/202348662},
	archivePrefix = {arXiv},
	eprint = {2403.05143},
	primaryClass = {astro-ph.GA},
	adsurl = {https://ui.adsabs.harvard.edu/abs/2024A&A...686A..42H}
}

@ARTICLE{Culpa2022,
	author = {{Culpan}, R. and {Geier}, S. and {Reindl}, N. and {Pelisoli}, I. and {Gentile Fusillo}, N. and {Vorontseva}, A.},
	title = "{The population of hot subdwarf stars studied with Gaia. IV. Catalogues of hot subluminous stars based on Gaia EDR3}",
	journal = {\aap},
	year = 2022,
	month = jun,
	volume = {662},
	eid = {A40},
	pages = {A40},
	doi = {10.1051/0004-6361/202243337},
	archivePrefix = {arXiv},
	eprint = {2203.07938},
	primaryClass = {astro-ph.SR},
	adsurl = {https://ui.adsabs.harvard.edu/abs/2022A&A...662A..40C}
}

@ARTICLE{OGrady2020,
	author = {{O'Grady}, Anna J.~G. and {Drout}, Maria R. and {Shappee}, B.~J. and {Bauer}, Evan B. and {Fuller}, Jim and {Kochanek}, C.~S. and {Jayasinghe}, T. and {Gaensler}, B.~M. and {Stanek}, K.~Z. and {Holoien}, Thomas W. -S. and {Prieto}, J.~L. and {Thompson}, Todd A.},
	title = "{Cool, Luminous, and Highly Variable Stars in the Magellanic Clouds from ASAS-SN: Implications for Thorne-{\.Z}ytkow Objects and Super-asymptotic Giant Branch Stars}",
	journal = {\apj},
	year = 2020,
	month = oct,
	volume = {901},
	number = {2},
	eid = {135},
	pages = {135},
	doi = {10.3847/1538-4357/abafad},
	archivePrefix = {arXiv},
	eprint = {2008.06563},
	primaryClass = {astro-ph.SR},
	adsurl = {https://ui.adsabs.harvard.edu/abs/2020ApJ...901..135O}
}

@ARTICLE{Grondin2024arXiv240704775G,
	author = {{Grondin}, Steffani M. and {Drout}, Maria R. and {Nordhaus}, Jason and {Muirhead}, Philip S. and {Speagle}, Joshua S. and {Chornock}, Ryan},
	title = "{The First Catalog of Candidate White Dwarf─Main-sequence Binaries in Open Star Clusters: A New Window into Common Envelope Evolution}",
	journal = {\apj},
	year = 2024,
	month = nov,
	volume = {976},
	number = {1},
	eid = {102},
	pages = {102},
	doi = {10.3847/1538-4357/ad7500},
	archivePrefix = {arXiv},
	eprint = {2407.04775},
	primaryClass = {astro-ph.SR},
	adsurl = {https://ui.adsabs.harvard.edu/abs/2024ApJ...976..102G}
}

@ARTICLE{Yan2026,
	author = {{Yan}, Huahui and {Wang}, Li and {Miller}, David R. and {He}, Chenyu and {Lin}, Jiamao and {Pang}, Xiaoying and {Zhao}, Jingkun and {Guo}, Jincheng and {de Grijs}, Richard and {Ge}, Hongwei and et al.},
	title = "{Searching for White Dwarf Candidates Formed Through Binary Evolution in Star Clusters}",
	journal = {\apj},
	year = 2026,
	month = mar,
	volume = {999},
	number = {2},
	eid = {167},
	pages = {167},
	doi = {10.3847/1538-4357/ae3f07},
	archivePrefix = {arXiv},
	eprint = {2602.16550},
	primaryClass = {astro-ph.SR},
	adsurl = {https://ui.adsabs.harvard.edu/abs/2026ApJ...999..167Y}
}

@ARTICLE{Tarricq2021,
	author = {{Tarricq}, Y. and {Soubiran}, C. and {Casamiquela}, L. and {Cantat-Gaudin}, T. and {Chemin}, L. and {Anders}, F. and {Antoja}, T. and {Romero-G{\'o}mez}, M. and {Figueras}, F. and {Jordi}, C. and {Bragaglia}, A. and {Balaguer-N{\'u}{\~n}ez}, L. and {Carrera}, R. and {Castro-Ginard}, A. and {Moitinho}, A. and {Ramos}, P. and {Bossini}, D.},
	title = "{3D kinematics and age distribution of the open cluster population}",
	journal = {\aap},
	year = 2021,
	month = mar,
	volume = {647},
	eid = {A19},
	pages = {A19},
	doi = {10.1051/0004-6361/202039388},
	archivePrefix = {arXiv},
	eprint = {2012.04017},
	primaryClass = {astro-ph.GA},
	adsurl = {https://ui.adsabs.harvard.edu/abs/2021A&A...647A..19T}
}

@ARTICLE{Jadhav2024,
	author = {{Jadhav}, Vikrant V. and {Kroupa}, Pavel and {Wu}, Wenjie and {Pflamm-Altenburg}, Jan and {Thies}, Ingo},
	title = "{The spin, expansion, and contraction of open star clusters}",
	journal = {\aap},
	year = 2024,
	month = jul,
	volume = {687},
	eid = {A89},
	pages = {A89},
	doi = {10.1051/0004-6361/202349115},
	archivePrefix = {arXiv},
	eprint = {2404.05327},
	primaryClass = {astro-ph.GA},
	adsurl = {https://ui.adsabs.harvard.edu/abs/2024A&A...687A..89J}
}

@ARTICLE{Angelo2025,
	author = {{Angelo}, M.~S. and {Santos}, J.~F.~C. and {Corradi}, W.~J.~B. and {Maia}, F.~F.~S.},
	title = "{Exploring the dynamical state of Galactic open clusters using Gaia DR3 and observational parameters}",
	journal = {\mnras},
	year = 2025,
	month = may,
	volume = {539},
	number = {3},
	pages = {2513-2536},
	doi = {10.1093/mnras/staf584},
	archivePrefix = {arXiv},
	eprint = {2504.06362},
	primaryClass = {astro-ph.GA},
	adsurl = {https://ui.adsabs.harvard.edu/abs/2025MNRAS.539.2513A}
}

@ARTICLE{Dias2021,
	author = {{Dias}, W.~S. and {Monteiro}, H. and {Moitinho}, A. and {L{\'e}pine}, J.~R.~D. and {Carraro}, G. and {Paunzen}, E. and {Alessi}, B. and {Villela}, L.},
	title = "{Updated parameters of 1743 open clusters based on Gaia DR2}",
	journal = {\mnras},
	year = 2021,
	month = jun,
	volume = {504},
	number = {1},
	pages = {356-371},
	doi = {10.1093/mnras/stab770},
	archivePrefix = {arXiv},
	eprint = {2103.12829},
	primaryClass = {astro-ph.SR},
	adsurl = {https://ui.adsabs.harvard.edu/abs/2021MNRAS.504..356D}
}

@ARTICLE{Miller2026,
	author = {{Miller}, David R. and {Caiazzo}, Ilaria and {Heyl}, Jeremy and {Richer}, Harvey B. and {Hollands}, Mark A. and {Tremblay}, Pier-Emmanuel and {El-Badry}, Kareem and {Rodriguez}, Antonio C. and {Vanderbosch}, Zachary P.},
	title = "{The White Dwarf Initial─Final Mass Relation from Open Clusters in Gaia DR3}",
	journal = {\apj},
	year = 2026,
	month = jan,
	volume = {996},
	number = {1},
	eid = {69},
	pages = {69},
	doi = {10.3847/1538-4357/ae18c8},
	archivePrefix = {arXiv},
	eprint = {2510.24877},
	primaryClass = {astro-ph.SR},
	adsurl = {https://ui.adsabs.harvard.edu/abs/2026ApJ...996...69M}
}

@ARTICLE{Bressan2012,
	author = {{Bressan}, Alessandro and {Marigo}, Paola and {Girardi}, L{\'e}o. and {Salasnich}, Bernardo and {Dal Cero}, Claudia and {Rubele}, Stefano and {Nanni}, Ambra},
	title = "{PARSEC: stellar tracks and isochrones with the PAdova and TRieste Stellar Evolution Code}",
	journal = {\mnras},
	year = 2012,
	month = nov,
	volume = {427},
	number = {1},
	pages = {127-145},
	doi = {10.1111/j.1365-2966.2012.21948.x},
	archivePrefix = {arXiv},
	eprint = {1208.4498},
	primaryClass = {astro-ph.SR},
	adsurl = {https://ui.adsabs.harvard.edu/abs/2012MNRAS.427..127B}
}

@ARTICLE{Nguyen2022,
	author = {{Nguyen}, C.~T. and {Costa}, G. and {Girardi}, L. and {Volpato}, G. and {Bressan}, A. and {Chen}, Y. and {Marigo}, P. and {Fu}, X. and {Goudfrooij}, P.},
	title = "{PARSEC V2.0: Stellar tracks and isochrones of low- and intermediate-mass stars with rotation}",
	journal = {\aap},
	year = 2022,
	month = sep,
	volume = {665},
	eid = {A126},
	pages = {A126},
	doi = {10.1051/0004-6361/202244166},
	archivePrefix = {arXiv},
	eprint = {2207.08642},
	primaryClass = {astro-ph.SR},
	adsurl = {https://ui.adsabs.harvard.edu/abs/2022A&A...665A.126N}
}

@ARTICLE{CantatGaudin2020,
	author = {{Cantat-Gaudin}, T. and {Anders}, F. and {Castro-Ginard}, A. and {Jordi}, C. and {Romero-G{\'o}mez}, M. and {Soubiran}, C. and {Casamiquela}, L. and {Tarricq}, Y. and {Moitinho}, A. and {Vallenari}, A. and {Bragaglia}, A. and {Krone-Martins}, A. and {Kounkel}, M.},
	title = "{Painting a portrait of the Galactic disc with its stellar clusters}",
	journal = {\aap},
	year = 2020,
	month = aug,
	volume = {640},
	eid = {A1},
	pages = {A1},
	doi = {10.1051/0004-6361/202038192},
	archivePrefix = {arXiv},
	eprint = {2004.07274},
	primaryClass = {astro-ph.GA},
	adsurl = {https://ui.adsabs.harvard.edu/abs/2020A&A...640A...1C}
}

@ARTICLE{Bellm2019,
	author = {{Bellm}, Eric C. and {Kulkarni}, Shrinivas R. and {Graham}, Matthew J. and {Dekany}, Richard and {Smith}, Roger M. and {Riddle}, Reed and {Masci}, Frank J. and {Helou}, George and {Prince}, Thomas A. and {Adams}, Scott M. and {Barbarino}, C. and {Barlow}, Tom and {Bauer}, James and {Beck}, Ron and {Belicki}, Justin and {Biswas}, Rahul and {Blagorodnova}, Nadejda and {Bodewits}, Dennis and {Bolin}, Bryce and {Brinnel}, Valery and {Brooke}, Tim and {Bue}, Brian and {Bulla}, Mattia and {Burruss}, Rick and {Cenko}, S. Bradley and {Chang}, Chan-Kao and {Connolly}, Andrew and {Coughlin}, Michael and {Cromer}, John and {Cunningham}, Virginia and {De}, Kishalay and {Delacroix}, Alex and {Desai}, Vandana and {Duev}, Dmitry A. and {Eadie}, Gwendolyn and {Farnham}, Tony L. and {Feeney}, Michael and {Feindt}, Ulrich and {Flynn}, David and {Franckowiak}, Anna and {Frederick}, S. and {Fremling}, C. and {Gal-Yam}, Avishay and {Gezari}, Suvi and {Giomi}, Matteo and {Goldstein}, Daniel A. and {Golkhou}, V. Zach and {Goobar}, Ariel and {Groom}, Steven and {Hacopians}, Eugean and {Hale}, David and {Henning}, John and {Ho}, Anna Y.~Q. and {Hover}, David and {Howell}, Justin and {Hung}, Tiara and {Huppenkothen}, Daniela and {Imel}, David and {Ip}, Wing-Huen and {Ivezi{\'c}}, {\v{Z}}eljko and {Jackson}, Edward and {Jones}, Lynne and {Juric}, Mario and {Kasliwal}, Mansi M. and {Kaspi}, S. and {Kaye}, Stephen and {Kelley}, Michael S.~P. and {Kowalski}, Marek and {Kramer}, Emily and {Kupfer}, Thomas and {Landry}, Walter and {Laher}, Russ R. and {Lee}, Chien-De and {Lin}, Hsing Wen and {Lin}, Zhong-Yi and {Lunnan}, Ragnhild and {Giomi}, Matteo and {Mahabal}, Ashish and {Mao}, Peter and {Miller}, Adam A. and {Monkewitz}, Serge and {Murphy}, Patrick and {Ngeow}, Chow-Choong and {Nordin}, Jakob and {Nugent}, Peter and {Ofek}, Eran and {Patterson}, Maria T. and {Penprase}, Bryan and {Porter}, Michael and {Rauch}, Ludwig and {Rebbapragada}, Umaa and {Reiley}, Dan and {Rigault}, Mickael and {Rodriguez}, Hector and {van Roestel}, Jan and {Rusholme}, Ben and {van Santen}, Jakob and {Schulze}, S. and {Shupe}, David L. and {Singer}, Leo P. and {Soumagnac}, Maayane T. and {Stein}, Robert and {Surace}, Jason and {Sollerman}, Jesper and {Szkody}, Paula and {Taddia}, F. and {Terek}, Scott and {Van Sistine}, Angela and {van Velzen}, Sjoert and {Vestrand}, W. Thomas and {Walters}, Richard and {Ward}, Charlotte and {Ye}, Quan-Zhi and {Yu}, Po-Chieh and {Yan}, Lin and {Zolkower}, Jeffry},
	title = "{The Zwicky Transient Facility: System Overview, Performance, and First Results}",
	journal = {\pasp},
	year = 2019,
	month = jan,
	volume = {131},
	number = {995},
	pages = {018002},
	doi = {10.1088/1538-3873/aaecbe},
	archivePrefix = {arXiv},
	eprint = {1902.01932},
	primaryClass = {astro-ph.IM},
	adsurl = {https://ui.adsabs.harvard.edu/abs/2019PASP..131a8002B}
}

@ARTICLE{VanderPlas2018,
	author = {{VanderPlas}, Jacob T.},
	title = "{Understanding the Lomb-Scargle Periodogram}",
	journal = {\apjs},
	year = 2018,
	month = may,
	volume = {236},
	number = {1},
	eid = {16},
	pages = {16},
	doi = {10.3847/1538-4365/aab766},
	archivePrefix = {arXiv},
	eprint = {1703.09824},
	primaryClass = {astro-ph.IM},
	adsurl = {https://ui.adsabs.harvard.edu/abs/2018ApJS..236...16V}
}

@ARTICLE{Kupfer2019,
	author = {{Kupfer}, Thomas and {Bauer}, Evan B. and {Burdge}, Kevin B. and {Bellm}, Eric C. and {Bildsten}, Lars and {Fuller}, Jim and {Hermes}, JJ and {Kulkarni}, Shrinivas R. and {Prince}, Thomas A. and {van Roestel}, Jan and {Dekany}, Richard and {Duev}, Dmitry A. and {Feeney}, Michael and {Giomi}, Matteo and {Graham}, Matthew J. and {Kaye}, Stephen and {Laher}, Russ R. and {Masci}, Frank J. and {Porter}, Michael and {Riddle}, Reed and {Shupe}, David L. and {Smith}, Roger M. and {Soumagnac}, Maayane T. and {Szkody}, Paula and {Ward}, Charlotte},
	title = "{A New Class of Large-amplitude Radial-mode Hot Subdwarf Pulsators}",
	journal = {\apjl},
	year = 2019,
	month = jun,
	volume = {878},
	number = {2},
	eid = {L35},
	pages = {L35},
	doi = {10.3847/2041-8213/ab263c},
	archivePrefix = {arXiv},
	eprint = {1906.00979},
	primaryClass = {astro-ph.GA},
	adsurl = {https://ui.adsabs.harvard.edu/abs/2019ApJ...878L..35K}
}

@ARTICLE{Kupfer2021,
	author = {{Kupfer}, Thomas and {Prince}, Thomas A. and {van Roestel}, Jan and {Bellm}, Eric C. and {Bildsten}, Lars and {Coughlin}, Michael W. and {Drake}, Andrew J. and {Graham}, Matthew J. and {Klein}, Courtney and {Kulkarni}, Shrinivas R. and {Masci}, Frank J. and {Walters}, Richard and {Andreoni}, Igor and {Biswas}, Rahul and {Bradshaw}, Corey and {Duev}, Dmitry A. and {Dekany}, Richard and {Guidry}, Joseph A. and {Hermes}, J.~J. and {Laher}, Russ R. and {Riddle}, Reed},
	title = "{Year 1 of the ZTF high-cadence Galactic plane survey: strategy, goals, and early results on new single-mode hot subdwarf B-star pulsators}",
	journal = {\mnras},
	year = 2021,
	month = jul,
	volume = {505},
	number = {1},
	pages = {1254-1267},
	doi = {10.1093/mnras/stab1344},
	archivePrefix = {arXiv},
	eprint = {2105.02758},
	primaryClass = {astro-ph.SR},
	adsurl = {https://ui.adsabs.harvard.edu/abs/2021MNRAS.505.1254K}
}

@ARTICLE{Hook2004,
	author = {{Hook}, I.~M. and {J{\o}rgensen}, Inger and {Allington-Smith}, J.~R. and {Davies}, R.~L. and {Metcalfe}, N. and {Murowinski}, R.~G. and {Crampton}, D.},
	title = "{The Gemini-North Multi-Object Spectrograph: Performance in Imaging, Long-Slit, and Multi-Object Spectroscopic Modes}",
	journal = {\pasp},
	year = 2004,
	month = may,
	volume = {116},
	number = {819},
	pages = {425-440},
	doi = {10.1086/383624},
	adsurl = {https://ui.adsabs.harvard.edu/abs/2004PASP..116..425H}
}

@ARTICLE{Peter2012,
	author = {{N{\'e}meth}, P{\'e}ter and {Kawka}, Ad{\'e}la and {Vennes}, St{\'e}phane},
	title = "{A selection of hot subluminous stars in the GALEX survey - II. Subdwarf atmospheric parameters}",
	journal = {\mnras},
	year = 2012,
	month = dec,
	volume = {427},
	number = {3},
	pages = {2180-2211},
	doi = {10.1111/j.1365-2966.2012.22009.x},
	archivePrefix = {arXiv},
	eprint = {1211.0323},
	primaryClass = {astro-ph.SR},
	adsurl = {https://ui.adsabs.harvard.edu/abs/2012MNRAS.427.2180N}
}

@ARTICLE{Hubeny1995,
	author = {{Hubeny}, I. and {Lanz}, T.},
	title = "{Non-LTE Line-blanketed Model Atmospheres of Hot Stars. I. Hybrid Complete Linearization/Accelerated Lambda Iteration Method}",
	journal = {\apj},
	year = 1995,
	month = feb,
	volume = {439},
	pages = {875},
	doi = {10.1086/175226},
	adsurl = {https://ui.adsabs.harvard.edu/abs/1995ApJ...439..875H}
}

@ARTICLE{Lanz2003,
	author = {{Lanz}, Thierry and {Hubeny}, Ivan},
	title = "{A Grid of Non-LTE Line-blanketed Model Atmospheres of O-Type Stars}",
	journal = {\apjs},
	year = 2003,
	month = jun,
	volume = {146},
	number = {2},
	pages = {417-441},
	doi = {10.1086/374373},
	archivePrefix = {arXiv},
	eprint = {astro-ph/0210157},
	primaryClass = {astro-ph},
	adsurl = {https://ui.adsabs.harvard.edu/abs/2003ApJS..146..417L}
}

@ARTICLE{Bradshaw2026,
	author = {{Bradshaw}, Corey W. and {Kupfer}, Thomas and {Kosakowski}, Alekzander R. and {Barlow}, Brad N. and {Dorsch}, Matti},
	title = "{Spectroscopic follow-up of hot subdwarf variables found in ZTF: Atmospheric and fundamental properties of radial-mode sdB pulsators}",
	journal = {\aap},
	year = 2026,
	month = apr,
	volume = {708},
	eid = {A197},
	pages = {A197},
	doi = {10.1051/0004-6361/202557985},
	archivePrefix = {arXiv},
	eprint = {2603.00240},
	primaryClass = {astro-ph.SR},
	adsurl = {https://ui.adsabs.harvard.edu/abs/2026A&A...708A.197B}
}

@ARTICLE{Han2008,
	author = {{Han}, Z.},
	title = "{A possible solution for the lack of EHB binaries in globular clusters}",
	journal = {\aap},
	year = 2008,
	month = jun,
	volume = {484},
	number = {3},
	pages = {L31-L34},
	doi = {10.1051/0004-6361:200809614},
	archivePrefix = {arXiv},
	eprint = {0804.4535},
	primaryClass = {astro-ph},
	adsurl = {https://ui.adsabs.harvard.edu/abs/2008A&A...484L..31H}
}

@ARTICLE{Han2002,
	author = {{Han}, Z. and {Podsiadlowski}, Ph. and {Maxted}, P.~F.~L. and {Marsh}, T.~R. and {Ivanova}, N.},
	title = "{The origin of subdwarf B stars - I. The formation channels}",
	journal = {\mnras},
	year = 2002,
	month = oct,
	volume = {336},
	number = {2},
	pages = {449-466},
	doi = {10.1046/j.1365-8711.2002.05752.x},
	archivePrefix = {arXiv},
	eprint = {astro-ph/0206130},
	primaryClass = {astro-ph},
	adsurl = {https://ui.adsabs.harvard.edu/abs/2002MNRAS.336..449H}
}

@ARTICLE{Han2003,
	author = {{Han}, Z. and {Podsiadlowski}, Ph. and {Maxted}, P.~F.~L. and {Marsh}, T.~R.},
	title = "{The origin of subdwarf B stars - II}",
	journal = {\mnras},
	year = 2003,
	month = may,
	volume = {341},
	number = {2},
	pages = {669-691},
	doi = {10.1046/j.1365-8711.2003.06451.x},
	archivePrefix = {arXiv},
	eprint = {astro-ph/0301380},
	primaryClass = {astro-ph},
	adsurl = {https://ui.adsabs.harvard.edu/abs/2003MNRAS.341..669H}
}

@ARTICLE{Paxton2011,
	author = {{Paxton}, B. and {Bildsten}, L. and {Dotter}, A. and {Herwig}, F. and {Lesaffre}, P. and {Timmes}, F.},
	title = {{Modules for Experiments in Stellar Astrophysics (MESA)}},
	journal = {\apjs},
	archivePrefix = {arXiv},
	eprint = {1009.1622},
	primaryClass = {astro-ph.SR},
	year = {2011},
	month = {jan},
	volume = {192},
	eid = {3},
	pages = {3},
	doi = {10.1088/0067-0049/192/1/3},
	adsurl = {https://ui.adsabs.harvard.edu/abs/2011ApJS..192....3P},
}

@ARTICLE{Paxton2013,
	author = {{Paxton}, B. and {Cantiello}, M. and {Arras}, P. and {Bildsten}, L. and {Brown}, E.~F. and {Dotter}, A. and {Mankovich}, C. and {Montgomery}, M.~H. and {Stello}, D. and {Timmes}, F.~X. and {Townsend}, R.},
	title = {{Modules for Experiments in Stellar Astrophysics (MESA): Planets, Oscillations, Rotation, and Massive Stars}},
	journal = {\apjs},
	archivePrefix = {arXiv},
	eprint = {1301.0319},
	primaryClass = {astro-ph.SR},
	year = {2013},
	month = {sep},
	volume = {208},
	eid = {4},
	pages = {4},
	doi = {10.1088/0067-0049/208/1/4},
	adsurl = {https://ui.adsabs.harvard.edu/abs/2013ApJS..208....4P},
}

@ARTICLE{Paxton2015,
	author = {{Paxton}, B. and {Marchant}, P. and {Schwab}, J. and {Bauer}, E.~B. and {Bildsten}, L. and {Cantiello}, M. and {Dessart}, L. and {Farmer}, R. and {Hu}, H. and {Langer}, N. and {Townsend}, R.~H.~D. and {Townsley}, D.~M. and {Timmes}, F.~X.},
	title = {{Modules for Experiments in Stellar Astrophysics (MESA): Binaries, Pulsations, and Explosions}},
	journal = {\apjs},
	archivePrefix = {arXiv},
	eprint = {1506.03146},
	primaryClass = {astro-ph.SR},
	year = {2015},
	month = {sep},
	volume = {220},
	eid = {15},
	pages = {15},
	doi = {10.1088/0067-0049/220/1/15},
	adsurl = {https://ui.adsabs.harvard.edu/abs/2015ApJS..220...15P},
}

@ARTICLE{Paxton2018,
	author = {{Paxton}, B. and {Schwab}, J. and {Bauer}, E.~B. and {Bildsten}, L. and {Blinnikov}, S. and {Duffell}, P. and {Farmer}, R. and {Goldberg}, J.~A. and {Marchant}, P. and {Sorokina}, E. and {Thoul}, A. and {Townsend}, R.~H.~D. and {Timmes}, F.~X.},
	title = {{Modules for Experiments in Stellar Astrophysics (MESA): Convective Boundaries, Element Diffusion, and Massive Star Explosions}},
	journal = {\apjs},
	archivePrefix = {arXiv},
	eprint = {1710.08424},
	primaryClass = {astro-ph.SR},
	year = {2018},
	month = {feb},
	volume = {234},
	eid = {34},
	pages = {34},
	doi = {10.3847/1538-4365/aaa5a8},
	adsurl = {https://ui.adsabs.harvard.edu/abs/2018ApJS..234...34P},
}

@ARTICLE{Paxton2019,
	author = {{Paxton}, Bill and {Smolec}, R. and {Schwab}, Josiah and {Gautschy}, A. and
	{Bildsten}, Lars and {Cantiello}, Matteo and {Dotter}, Aaron and
	{Farmer}, R. and {Goldberg}, Jared A. and {Jermyn}, Adam S. and
	{Kanbur}, S.~M. and {Marchant}, Pablo and {Thoul}, Anne and
	{Townsend}, Richard H.~D. and {Wolf}, William M. and {Zhang}, Michael and
	{Timmes}, F.~X.},
	title = "{Modules for Experiments in Stellar Astrophysics (MESA): Pulsating Variable Stars, Rotation, Convective Boundaries, and Energy Conservation}",
	journal = {\apjs},
	year = "2019",
	month = "Jul",
	volume = {243},
	number = {1},
	eid = {10},
	pages = {10},
	doi = {10.3847/1538-4365/ab2241},
	archivePrefix = {arXiv},
	eprint = {1903.01426},
	primaryClass = {astro-ph.SR},
	adsurl = {https://ui.adsabs.harvard.edu/abs/2019ApJS..243...10P}
}

@ARTICLE{Jermyn2023,
	author = {{Jermyn}, Adam S. and {Bauer}, Evan B. and {Schwab}, Josiah and {Farmer}, R. and {Ball}, Warrick H. and {Bellinger}, Earl P. and {Dotter}, Aaron and {Joyce}, Meridith and {Marchant}, Pablo and {Mombarg}, Joey S.~G. and {Wolf}, William M. and {Sunny Wong}, Tin Long and {Cinquegrana}, Giulia C. and {Farrell}, Eoin and {Smolec}, R. and {Thoul}, Anne and {Cantiello}, Matteo and {Herwig}, Falk and {Toloza}, Odette and {Bildsten}, Lars and {Townsend}, Richard H.~D. and {Timmes}, F.~X.},
	title = "{Modules for Experiments in Stellar Astrophysics (MESA): Time-dependent Convection, Energy Conservation, Automatic Differentiation, and Infrastructure}",
	journal = {\apjs},
	year = 2023,
	month = mar,
	volume = {265},
	number = {1},
	eid = {15},
	pages = {15},
	doi = {10.3847/1538-4365/acae8d},
	archivePrefix = {arXiv},
	eprint = {2208.03651},
	primaryClass = {astro-ph.SR},
	adsurl = {https://ui.adsabs.harvard.edu/abs/2023ApJS..265...15J}
}

@ARTICLE{Meng2020,
	author = {{Meng}, Xiang-Cun and {Han}, Zhan-Wen and {Podsiadlowski}, Philipp and {Li}, Jiao},
	title = "{Blue Large-amplitude Pulsators: The Possible Surviving Companions of Type Ia Supernovae}",
	journal = {\apj},
	year = 2020,
	month = nov,
	volume = {903},
	number = {2},
	pages = {100},
	doi = {10.3847/1538-4357/abbb8e}
}

@ARTICLE{Meng2021,
	author = {{Meng}, Xiang-Cun and {Luo}, Yang-Ping},
	title = "{Hot subdwarfs from the surviving companions of the white dwarf + main-sequence channel of Type Ia supernovae}",
	journal = {\mnras},
	year = 2021,
	month = nov,
	volume = {507},
	number = {3},
	pages = {4603-4617},
	doi = {10.1093/mnras/stab2369}
}

@ARTICLE{Ge2015,
	author = {{Ge}, Hongwei and {Webbink}, Ronald F. and {Chen}, Xuefei and {Han}, Zhanwen},
	title = "{Adiabatic Mass Loss in Binary Stars. II. From Zero-age Main Sequence to the Base of the Giant Branch}",
	journal = {\apj},
	year = 2015,
	month = oct,
	volume = {812},
	number = {1},
	eid = {40},
	pages = {40},
	doi = {10.1088/0004-637X/812/1/40},
	archivePrefix = {arXiv},
	eprint = {1507.04843},
	primaryClass = {astro-ph.SR},
	adsurl = {https://ui.adsabs.harvard.edu/abs/2015ApJ...812...40G}
}

@ARTICLE{Clausen2012,
	author = {{Clausen}, Drew and {Wade}, Richard A. and {Kopparapu}, Ravi Kumar and {O'Shaughnessy}, Richard},
	title = "{Population Synthesis of Hot Subdwarfs: A Parameter Study}",
	journal = {\apj},
	year = 2012,
	month = feb,
	volume = {746},
	number = {2},
	eid = {186},
	pages = {186},
	doi = {10.1088/0004-637X/746/2/186},
	archivePrefix = {arXiv},
	eprint = {1201.0012},
	primaryClass = {astro-ph.SR},
	adsurl = {https://ui.adsabs.harvard.edu/abs/2012ApJ...746..186C}
}

@ARTICLE{Gies1998,
	author = {{Gies}, Douglas R. and {Bagnuolo}, Jr., William G. and {Ferrara}, Elizabeth C. and {Kaye}, Anthony B. and {Thaller}, Michelle L. and {Penny}, Laura R. and {Peters}, Geraldine J.},
	title = "{Hubble Space Telescope Goddard High Resolution Spectrograph Observations of the Be + sdO Binary {\ensuremath{\varphi}} Persei}",
	journal = {\apj},
	year = 1998,
	month = jan,
	volume = {493},
	number = {1},
	pages = {440-450},
	doi = {10.1086/305113},
	adsurl = {https://ui.adsabs.harvard.edu/abs/1998ApJ...493..440G}
}

@ARTICLE{Mourard2015,
	author = {{Mourard}, D. and {Monnier}, J.~D. and {Meilland}, A. and {Gies}, D. and {Millour}, F. and {Benisty}, M. and {Che}, X. and {Grundstrom}, E.~D. and {Ligi}, R. and {Schaefer}, G. and {Baron}, F. and {Kraus}, S. and {Zhao}, M. and {Pedretti}, E. and {Berio}, P. and {Clausse}, J.~M. and {Nardetto}, N. and {Perraut}, K. and {Spang}, A. and {Stee}, P. and {Tallon-Bosc}, I. and {McAlister}, H. and {ten Brummelaar}, T. and {Ridgway}, S.~T. and {Sturmann}, J. and {Sturmann}, L. and {Turner}, N. and {Farrington}, C.},
	title = "{Spectral and spatial imaging of the Be+sdO binary <ASTROBJ>{\ensuremath{\phi}} Persei</ASTROBJ>}",
	journal = {\aap},
	year = 2015,
	month = may,
	volume = {577},
	eid = {A51},
	pages = {A51},
	doi = {10.1051/0004-6361/201425141},
	archivePrefix = {arXiv},
	eprint = {1503.03423},
	primaryClass = {astro-ph.SR},
	adsurl = {https://ui.adsabs.harvard.edu/abs/2015A&A...577A..51M}
}

@ARTICLE{Schootemeijer2018,
	author = {{Schootemeijer}, A. and {G{\"o}tberg}, Y. and {de Mink}, S.~E. and {Gies}, D. and {Zapartas}, E.},
	title = "{Clues about the scarcity of stripped-envelope stars from the evolutionary state of the sdO+Be binary system {\ensuremath{\varphi}} Persei}",
	journal = {\aap},
	year = 2018,
	month = jul,
	volume = {615},
	eid = {A30},
	pages = {A30},
	doi = {10.1051/0004-6361/201731194},
	archivePrefix = {arXiv},
	eprint = {1803.02379},
	primaryClass = {astro-ph.SR},
	adsurl = {https://ui.adsabs.harvard.edu/abs/2018A&A...615A..30S}
}

@ARTICLE{Baluev2008,
	author = {{Baluev}, R.~V.},
	title = "{Assessing the statistical significance of periodogram peaks}",
	journal = {\mnras},
	year = 2008,
	month = apr,
	volume = {385},
	number = {3},
	pages = {1279-1285},
	doi = {10.1111/j.1365-2966.2008.12689.x},
	archivePrefix = {arXiv},
	eprint = {0711.0330},
	primaryClass = {astro-ph},
	adsurl = {https://ui.adsabs.harvard.edu/abs/2008MNRAS.385.1279B}
}

@BOOK{Kippenhahn2013,
	author = {{Kippenhahn}, Rudolf and {Weigert}, Alfred and {Weiss}, Achim},
	title = "{Stellar Structure and Evolution}",
	year = 2013,
	doi = {10.1007/978-3-642-30304-3},
	adsurl = {https://ui.adsabs.harvard.edu/abs/2013sse..book.....K}
}

@ARTICLE{Baran2024,
	author = {{Baran}, A.~S. and {Charpinet}, S. and {{\O}stensen}, R.~H. and {Reed}, M.~D. and {Van Grootel}, V. and {Lyu}, C. and {Telting}, J.~H. and {N{\'e}meth}, P.},
	title = "{Short-period pulsating hot subdwarf stars observed by TESS. II. Northern ecliptic hemisphere}",
	journal = {\aap},
	year = 2024,
	month = jun,
	volume = {686},
	eid = {A65},
	pages = {A65},
	doi = {10.1051/0004-6361/202348571},
	archivePrefix = {arXiv},
	eprint = {2403.02384},
	primaryClass = {astro-ph.SR},
	adsurl = {https://ui.adsabs.harvard.edu/abs/2024A&A...686A..65B}
}

@ARTICLE{Uzundag2024,
	author = {{Uzundag}, Murat and {Krzesinski}, Jurek and {Pelisoli}, Ingrid and {N{\'e}meth}, P{\'e}ter and {Silvotti}, Roberto and {Vu{\v{c}}kovi{\'c}}, Maja and {Dawson}, Harry and {Geier}, Stephan},
	title = "{A comprehensive search for hot subdwarf stars using Gaia and TESS. I. Pulsating hot subdwarf B stars}",
	journal = {\aap},
	year = 2024,
	month = apr,
	volume = {684},
	eid = {A118},
	pages = {A118},
	doi = {10.1051/0004-6361/202348829},
	archivePrefix = {arXiv},
	eprint = {2401.17707},
	primaryClass = {astro-ph.SR},
	adsurl = {https://ui.adsabs.harvard.edu/abs/2024A&A...684A.118U}
}

@ARTICLE{Krzesinski2025,
	author = {{Krzesinski}, Jurek and {Uzundag}, Murat and {Kumari}, Goutham Anitha and {N{\'e}meth}, P{\'e}ter and {Pelisoli}, Ingrid and {Munday}, James and {Dawson}, Harry and {Vu{\v{c}}kovi{\'c}}, Maja and {Silvotti}, Roberto and {Geier}, Stephan and {Waniak}, Waclaw and {Siwak}, Michal},
	title = "{A comprehensive search for hot subdwarf stars using Gaia and TESS: II. Uncovering new pulsators and close binary systems}",
	journal = {\aap},
	year = 2025,
	month = aug,
	volume = {700},
	eid = {A71},
	pages = {A71},
	doi = {10.1051/0004-6361/202452606},
	adsurl = {https://ui.adsabs.harvard.edu/abs/2025A&A...700A..71K}
}
\bibliographystyle{aasjournalv7}



\end{document}